\documentclass[webpdf,contemporary,large]{oup-authoring-template}%

\graphicspath{{Fig/}}

\usepackage{graphicx}
\usepackage{balance}
\graphicspath{ {./figures/} }
\usepackage{subcaption}
\usepackage{amsmath}
\usepackage{lineno}
\usepackage{booktabs}
\usepackage{arydshln}
\usepackage{chngcntr}
\usepackage{xcolor}
\usepackage{xurl}
\usepackage{hyperref}

\setcitestyle{numbers}

\makeatletter
\newcommand\setcurrentname[1]{\def\@currentlabelname{#1}}
\makeatother

\theoremstyle{thmstyleone}%
\theoremstyle{thmstyletwo}%
\theoremstyle{thmstylethree}%

\begin{document}

\journaltitle{PNAS Nexus}
\DOI{DOI HERE}
\copyrightyear{2024}
\pubyear{2024}
\access{Advance Access Publication Date: Day Month Year}
\appnotes{Paper}

\firstpage{1}


\title{The Potential Impact of AI Innovations on U.S. Occupations}

\author[a]{Ali Akbar Septiandri}
\author[a]{Marios Constantinides}
\author[a,b$\ast$]{Daniele Quercia}

\authormark{Septiandri et al.}

\address[a]{\orgname{Nokia Bell Labs}, \orgaddress{\street{21 JJ Thomson Avenue}, \postcode{CB3 0FA}, \state{Cambridge}, \country{United Kingdom}}}

\address[b]{\orgname{King's College London}, \orgaddress{\street{Strand}, \postcode{WC2R 2LS}, \state{London}, \country{United Kingdom}}}

\corresp[$\ast$]{To whom correspondence should be addressed: \href{quercia@cantab.net}{quercia@cantab.net}}

\received{Date}{0}{Year}
\accepted{Date}{0}{Year}


\abstract{An occupation is comprised of interconnected tasks, and it is these tasks, not occupations themselves, that are affected by AI. To evaluate how tasks may be impacted, previous approaches utilized manual annotations or coarse-grained matching. Leveraging recent advancements in machine learning, we replace coarse-grained matching with more precise deep learning approaches. Introducing the AI Impact (AII) measure, we employ Deep Learning Natural Language Processing to automatically identify AI patents that may impact various occupational tasks at scale. Our methodology relies on a comprehensive dataset of 17,879 task descriptions and quantifies AI's potential impact through analysis of 24,758 AI patents filed with the United States Patent and Trademark Office (USPTO) between 2015 and 2022. Our results reveal that some occupations will potentially be impacted, and that impact is intricately linked to specific skills. These include not only routine tasks (codified as a series of steps), as previously thought, but also non-routine ones (e.g., diagnosing health conditions, programming computers, and tracking flight routes). However, 
AI's impact on labour is limited by the fact that some of the occupations affected are augmented rather than replaced (e.g., neurologists, software engineers,  air traffic controllers), and the sectors  affected are
experiencing labour shortages (e.g., IT, Healthcare, Transport).}


\boxedtext{
We introduce the AI Impact (AII) measure\footnote{Project page: \url{https://social-dynamics.net/aii}}, utilizing Deep Learning Natural Language Processing to automatically identify AI patents affecting occupational tasks. Our findings reveal that: 
\begin{enumerate}

\item AI's impact on occupations defies simple categorizations of task routineness. It intricately affects specific skills within tasks, from routine (e.g., scanning items) to non-routine (e.g., decision-making under stress by air traffic controllers), challenging the assumption that only routine tasks are susceptible.

\item AI's impact on labor may be limited by the fact that some of the affected occupations are augmented rather than replaced, and some of the sectors affected are experiencing labor shortages.
\end{enumerate}
}
\bigskip

\maketitle

\section*{Introduction}
\setcurrentname{Introduction}\phantomsection\label{sec:intro}

The rapid advancement of Artificial Intelligence (AI) has undeniably created new business opportunities~\cite{bonaventura2020predicting} but has also reshaped the labor market~\cite{webb2019impact, brynjolfsson2018machines, felten2021occupational, galaz2021artificial}, simultaneously reducing hiring in non-AI positions and altering the skill requirements of remaining job postings~\cite{acemoglu2022artificial}.  Exemplifying this phenomenon, the manufacturing sector has witnessed the automation of previously human-intensive assembly line tasks, while chatbots and virtual assistants have taken over routine inquiries and support functions in customer services \cite{kannan2019chatbot}. Recent AI advances, including generative AI, may further (re)shape occupations over the long term, fuelling growth in certain sectors and eroding others~\cite{ellingrud2023mckgenai}. AI automation has not only streamlined processes but has also generated economic benefits~\cite{acemoglu2020ai}, enabling companies to allocate resources more effectively and to redirect human capital towards higher-value, creative, and complex tasks---\emph{(re)skilling and upskilling} their workforce~\cite{jagannathan2019dominant}. However, this transformation has given rise to divergent viewpoints, with some scholars arguing for a future characterized by AI occupation displacement and mass unemployment~\cite{ford2015rise, west2018future}, while others posit that the AI revolution has the potential to enhance both productivity and quality of work~\cite{chui2017artificial}.

Previous literature on AI impact on occupations has primarily focused on two classes of methodologies. The first measures the impact of AI on occupations. More specifically, it breaks down occupations into a finite set of abilities (e.g., manual dexterity, persuasion) and measures the impact on those abilities. Two examples that illustrate this approach are the Frey and Osborne's~\cite{frey2017future} method and the AI Occupational Exposure (AIOE) method~\cite{felten2021occupational}. Frey and Osborne's method uses 9 abilities extracted from O*NET database, covering 70 occupations that were manually labeled and extended with a classifier to 702, including roles like clergy, dentists, and chief executives. In contrast, AIOE uses 52 abilities derived from the O*NET database, focusing on 10 Electronic Frontier Foundation (EFF) applications such as image and speech recognition, and language modeling. However, both of these methods share a common limitation. Their reliance on coarse-grained abilities in the computation of AI's impact may not fully capture the nuances of AI. Consider, for example, the ability of information ordering. Methods based on abilities may categorize tasks that involve organizing information in the same way, without distinguishing between the highly structured and complex information ordering required for database design and the simpler, routine information tasks of librarians such as alphabetizing files~\cite{coelli2019behind}.

The second class of methodologies measures the impact of AI on tasks rather than occupations. This concept is illustrated by Brynjolfsson \emph{et al.}'s Suitability for Machine Learning (SML) method~\cite{brynjolfsson2018machines}, which measures the impact of AI using a comprehensive set of 18,156 tasks spanning 964 different occupations. However, SML relies on the assessments of crowdworkers to determine the suitability of specific tasks for machine learning. This reliance may introduce subjective biases from annotators (e.g., varying levels of expertise or cultural factors may lead to inconsistencies), and poses challenges in terms of scalability. Also, similar to  Frey and Osborne's and AIOE, SML is limited by the static nature of its one-time manual labeling. As technology advances and new capabilities emerge such as Large Language Models (LLMs), relying solely on a fixed set of abilities or subjective assessments of task suitability for automation becomes increasingly inadequate. For example, a copywriter is likely to be impacted by LLMs~\cite{agrawal2022chatgpt}. However, if one were to examine copywriters at different points in time, such as in 2010 or 2015, the impact of AI on them would not be constant but would drastically change since language models were not as powerful back in 2010 as they are today. To fully capture the  impact of a fast moving technology such as AI, therefore, it is crucial for methods to be adaptable to the ever-evolving technological advancements. To gauge the likelihood of automation, it is essential to identify which systems are poised for construction and commercialization. Annual business surveys serve as a source for measuring the adoption of automation. However, they may be subject to biases  (e.g., respondents may over-report positive aspects, prioritize certain business operations, potentially neglect others, and interpret survey questions subjectively without standardized criteria) and are infrequently updated~\cite{acemoglu2022automation}. An alternative, more objective source is patents. Patents are a typical source in scholarly work to identify emerging technological innovations~\cite{straccamore2023urban, straccamore2023geography, arcaute2015constructing}. Prior work used patents to study the effect of automation and employment changes~\cite{mann2023benign, damioli2022supply}. By analyzing the text of U.S. patents granted between 1976 and 2014, Mann \emph{et al.}~\cite{mann2023benign} showed that the effect of automation differs across sectors. For example, the manufacturing industry, where most robots are used, experienced employment losses, while the service sector experienced employment gains; a finding that aligns with those reported by Autor and Dorn~\cite{autor2013growth}. More broadly, patents provide insights into emerging systems and technologies~\cite{chaturvedi2023automation}, leading Webb to study AI innovations by comparing occupational task descriptions with patent titles~\cite{webb2019impact}. This method employs a dictionary approach to identify verb-noun pairs associated with both tasks and patent titles. Another method, similar in its approach to Webb's term matching, employs a normalized term matching approach to determine the similarity between tasks and patents~\cite{montobbio2023labour}, and does so in the specific area of robotics rather than AI. The method most similar to ours was proposed in~\cite{autor2022new, kogan2023technology}, in which, using word embeddings, patents are matched with broad occupation categories from the American Community Survey. However, that level of categorization is not suitable for researching the characteristics of specific jobs. Overall, this second class of approaches has used either term matching or word embedding. The problem is that term matching does not capture the semantic meaning of words (e.g., it does not distinguish between `bank' as in `data bank' or as in a financial institution)~\cite{Jurafsky2008nlp}, and word embedding does not account for word ordering (e.g., `data entry and analysis' and `analysis of entry data' are considered similar based on word embeddings yet are two different tasks). As a result, these methods either miss relevant patents or return spurious task-patent matches, as detailed in Tables~\ref{tab:elevator} and~\ref{tab:task_patent_verb_noun}.

\begin{figure}[t!]
    \centering
    \includegraphics[width=1.\linewidth]{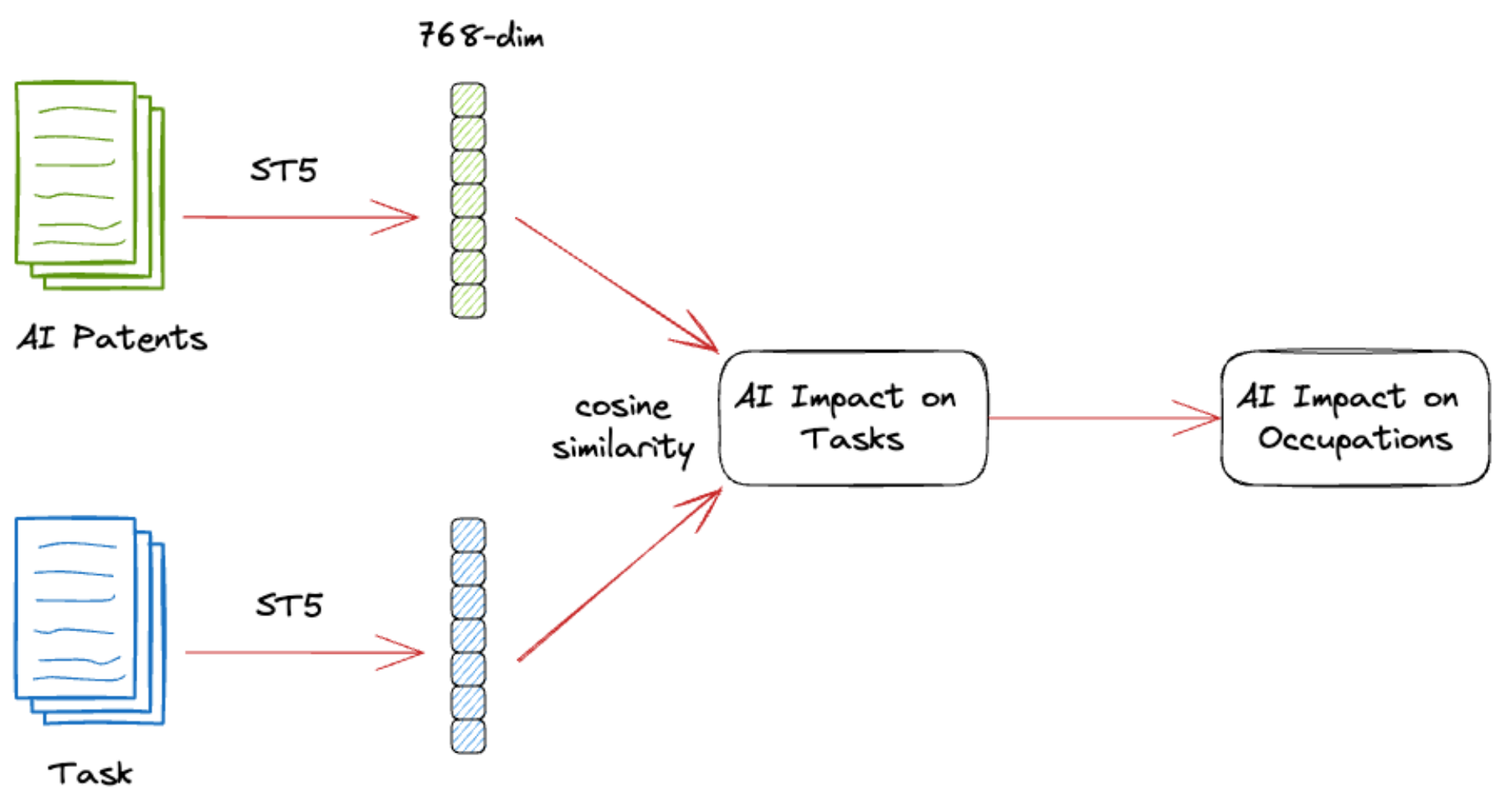}
    \caption{\textbf{Framework automating AI Impact (AII) measure.} Using the Sentence-T5 (ST5) model, we first generated two vector representations (embeddings): one capturing the semantic meaning of an AI patent, and the other the semantic meaning of a task description. We then computed the patent-task cosine similarity from the embeddings on all patent-task pairs. This process was conducted to identify which tasks were impacted by which AI patents. Finally, for each occupation, we calculated the proportion of impacted tasks out of the total ones, and this proportion determined the AII of that occupation.}
    \label{fig:schema}
\end{figure}

To overcome these limitations, we introduced and validated the AI Impact (AII) measure. AII utilizes 19,259 task descriptions from O*NET and assesses AI's potential impact through innovations found in 24,758 AI patents filed with the United States Patent and Trademark Office (USPTO) from 2015 to 2022. Built on Sentence-T5 (ST5)\cite{ni2021st5}, a natural language processing-focused deep learning framework (Figure \ref{fig:schema}), the method gauges semantic similarity between occupation task descriptions and patent descriptions (explained in ``\nameref{subsec:datasets}'') by embedding not individual words but the entire document (e.g., the entire patent's abstract), allowing for considering both semantic meaning and word ordering. The AII score is calculated in three steps. Firstly, the method identifies the most similar patent for each task based on maximum cosine similarity. Secondly, it categorizes a task as AI-impacted if its similarity with the most similar patent surpasses a threshold at the 90\textsuperscript{th} percentile, as previous literature suggested~\cite{chaturvedi2023automation} and this work further empirically validated (see ``\nameref{appendix:validation}'' in Supplementary Material). We select, for each task, the closest patent rather than counting the number of closely linked patents. This approach provides a more targeted understanding of the specific innovations or solutions directly relevant to that particular task, avoiding dilution of the analysis with potentially less pertinent or peripheral patents. Also, given that patents can be general, our approach addresses this by using the text not only in titles, as previous approaches did~\cite{webb2019impact},  but also in abstracts, which are more likely to contain application domains or references to specific tasks (illustrated in Table~\ref{tab:generic-patents}). Finally, the AII score for an occupation is computed by dividing the number of tasks impacted by AI patents by the total tasks for that occupation (as detailed in ``\nameref{subsec:new_method}''). For insights into AI's economic ramifications, the method aggregates AII scores for each occupation at industry sector-level.

\section*{Results}
\label{sec:results}

\subsection*{Validation with Historically Impacted Occupations}
To begin to understand the nature of the AII score, we first validated it empirically through two historical case studies: robots and software. We chose them  for three reasons. First, their introduction into the labor market has been associated with reductions in employment and wages~\cite{acemoglu2020robots,graetz2015robots}. Second, due to the recent emergence of these technologies, they are likely to provide insights into how the economy may respond to the introduction of AI. Lastly, these two historical cases have been used in previous works to empirically assess methodologies similar to ours~\cite{webb2019impact}, offering a basis for comparison.

\setlength{\dashlinedash}{0.2pt} 
\setlength{\dashlinegap}{1.5pt} 
\aboverulesep=0ex
\belowrulesep=0ex

\begin{table*}[t!]
    \centering
    \caption{20 most- and least-impacted occupations ranked by the AII (Artificial Intelligence Impact) measure.}
    \label{tab:job_aii}
    \scalebox{0.77}{
    \begin{tabular}{|l|l|l|}
    \hline
    \textbf{Rank} & \textbf{Most-impacted} & \textbf{Least-impacted} \\
    \hline
    1 & Cardiovascular Technologists and Technicians & Pile Driver Operators \\
    \hdashline
    2 & Sound Engineering Technicians & Dredge Operators \\
    \hdashline
    3 & Nuclear Medicine Technologists & Aircraft Cargo Handling Supervisors \\
    \hdashline
    4 & Air Traffic Controllers & Graders and Sorters, Agricultural Products \\
    \hdashline
    5 & Magnetic Resonance Imaging Technologists & Insurance Underwriters \\
    \hdashline
    6 & Electro-Mechanical and Mechatronics Technologists and Technicians & Floor Sanders and Finishers \\
    \hdashline
    7 & Orthodontists & Reinforcing Iron and Rebar Workers \\
    \hdashline
    8 & Power Distributors and Dispatchers & Farm Labor Contractors \\
    \hdashline
    9 & Neurologists & Administrative Services Managers \\
    \hdashline
    10 & Industrial Truck and Tractor Operators & Rock Splitters, Quarry \\
    \hdashline
    11 & Public Safety Telecommunicators & Brokerage Clerks \\
    \hdashline
    12 & Computer Numerically Controlled Tool Programmers & Podiatrists \\
    \hdashline
    13 & Security Guards & Helpers--Painters, Paperhangers, Plasterers, and Stucco Masons \\
    \hdashline
    14 & Remote Sensing Scientists and Technologists & Shipping, Receiving, and Inventory Clerks \\
    \hdashline
    15 & Machinists & Cooks, Short Order \\
    \hdashline
    16 & Radiologists & Team Assemblers \\
    \hdashline
    17 & Atmospheric and Space Scientists & Proofreaders and Copy Markers \\
    \hdashline
    18 & Computer Numerically Controlled Tool Operators & Butchers and Meat Cutters \\
    \hdashline
    19 & Textile Knitting and Weaving Machine Setters, Operators, and Tenders & Door-to-Door Sales Workers, News and Street Vendors, and Related Workers \\
    \hdashline
    20 & Medical Transcriptionists & Segmental Pavers \\
    \hline
    \end{tabular}
    }
\end{table*} 

We therefore adjusted the AII score to encompass exposure to robots and software rather than AI (as detailed in ``\nameref{appendix:wage}'' of Supplementary Material) by focusing on patents related to these two technologies. We studied  how robot and software exposure affected employment and wages using US census data from 1980 to 2010. Following Webb's methodology~\cite{webb2019impact}, which controlled for industry effects, educational levels, wage polarization, and off-shorability,  we found that introducing robots led to a 9\% decrease in employment and a 4\% decrease in wages, and that introducing software resulted in a 10\% decrease in employment and a 7\% decrease in wages during this period. These findings align with Webb's, indicating that occupations exposed to robots or software have decreased in number and pay lower wages. However, Webb's method, which relies on keyword matching, sometimes includes patents that should not be matched with certain job tasks (as detailed in ``\nameref{appendix:selected-occupations}'' in Supplementary Material), leading to larger decreases and an overestimation of AI's potential impact compared to our estimates.

\subsection*{Most- and Least-Impacted Occupations}
\label{subsec:impact_occupations}

We compared the potentially most-impacted (highest AII scores) occupations with the least-impacted (lowest AII scores) occupations (Table~\ref{tab:job_aii} only reports the 20 most- and least-impacted occupations for brevity and comparability with previous methods), and did so by thematically analyzing the AI patents associated with the tasks of each group's occupations (as described in ``\nameref{subsec:thematic}'').

\begin{figure}[t!]
    \centering
    \includegraphics[width=.4\textwidth]{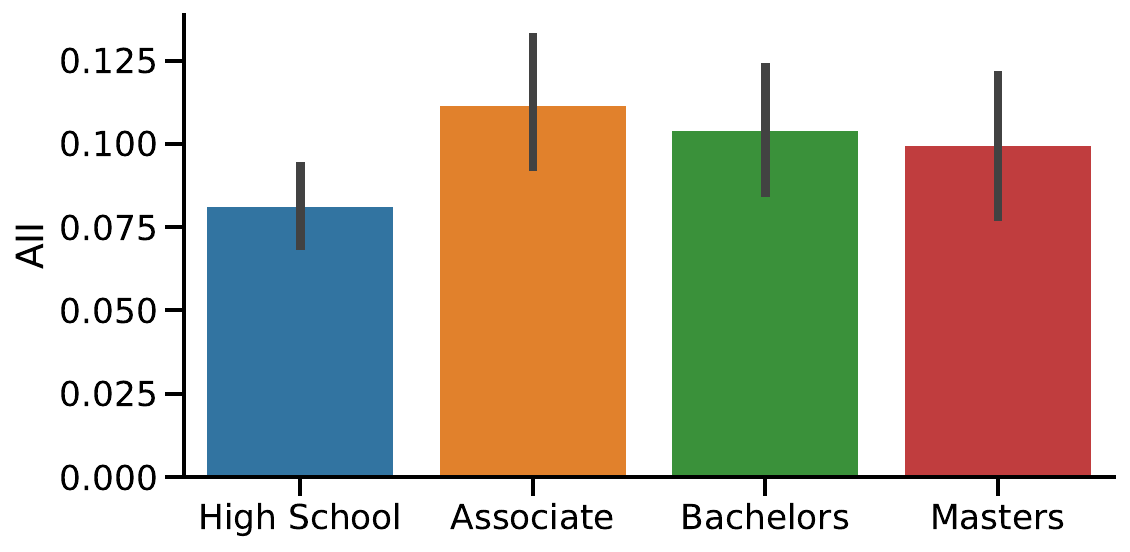}
    \caption{AII score binned by the level of education: high school, associate degrees from community colleges, bachelor's degrees, and master's degrees in science.  This binned score was obtained by averaging the scores across all occupations in a given education category, weighted by the total employment for those binned occupations.}
    \label{fig:education}
\end{figure}

\begin{table*}[t!]
    \centering
     \caption{The most-impacted occupations based on AII scores. Each occupation's score is calculated as the number of impacted tasks divided by the total number of tasks for that occupation. For each occupation, a task-patent pair is presented, corresponding to the most impacted task by the patent (i.e., the patent with the highest similarity score to the task) which is determined by calculating the textual similarity between the task's description and a patent's title plus abstract. }
    \label{tab:job_tasks_patents}
    \scalebox{0.7}{    
    \begin{tabular}{|c|p{4cm}p{2.1cm}p{6cm}p{4cm}ccc|c|}
        \hline
        \textbf{Rank} & \textbf{Occupation} & \textbf{Sector} & \textbf{Task} & \textbf{Patent} & \textbf{Similarity} & \textbf{\shortstack[c]{\# Impacted \\ Tasks}} & \textbf{\shortstack[c]{Total \\ Tasks}} & \textbf{AII} \\
        \hline
        1 & Cardiovascular Technologists and Technicians & Healthcare & Observe gauges, recorder, and video screens [...] during imaging of cardiovascular system. & Automated analysis of vasculature in coronary angiograms & 0.8753 & 16 & 25 & 0.64 \\
        \hdashline
        2 & Sound Engineering Technicians & Arts and entertainment & Record speech, music, and other sounds on recording media, using recording equipment. & Media capture and process system & 0.8293 & 8 & 14 & 0.57 \\
        \hdashline
        3 & Nuclear Medicine Technologists & Healthcare & Process cardiac function studies, using computer. & Electrocardiogram analysis & 0.8645 & 9 & 17 & 0.53 \\
        \hdashline
        4 & Air Traffic Controllers & Transportation &  Determine the timing or procedures for flight vector changes. & Constraint processing as an alternative to flight management systems & 0.8298 & 12 & 23 & 0.52 \\
        \hdashline
        5 & Magnetic Resonance Imaging Technologists & Healthcare & Operate optical systems to capture dynamic magnetic resonance imaging (MRI) images [...] & MRI system and method using neural network for detection of patient motion & 0.8615 & 12 & 23 & 0.52 \\
        \hdashline
        6 & Electro-Mechanical and Mechatronics Technologists and Technicians & Manufacturing & Train robots, using artificial intelligence software or interactive training techniques [...] & Backup control based continuous training of robots & 0.8846 & 18 & 35 & 0.51 \\
        \hdashline
        7 & Orthodontists & Healthcare & Study diagnostic records, such as medical or dental histories [...] to develop patient treatment plans. & Patient-Specific Therapy Planning Support Using Patient Matching & 0.8460 & 5 & 10 & 0.50 \\
        \hdashline
        8 & Power Distributors and Dispatchers & Utilities & Control, monitor, or operate equipment that regulates or distributes electricity or steam [...] & Power grid aware machine learning device & 0.8562 & 7 & 15 & 0.47 \\
        \hdashline
        9 & Neurologists & Healthcare & Interpret the results of neuroimaging studies, such as [...] Positron Emission Tomography (PET) scans. & Pet quantitative localization system and operation method thereof & 0.8352 & 11 & 24 & 0.46 \\
        \hdashline
        10 & Industrial Truck and Tractor Operators & Manufacturing & Move controls to drive gasoline- or electric-powered trucks, [...] & Autonomous Truck Unloading for Mining and Construction Applications & 0.8301 & 5 & 11 & 0.45 \\
        \hdashline
        11 & Public Safety Telecommunicators & Public administration & Test and adjust communication and alarm systems, and report malfunctions to maintenance units. & Security-Relevant Diagnostic Messages & 0.8237 & 8 & 18 & 0.44 \\
        \hdashline
        12 & Computer Numerically Controlled Tool Programmers & Manufacturing & Determine the sequence of machine operations, and select the proper cutting tools [...] & Methods and apparatuses for cutter path planning and for workpiece machining & 0.8376 & 7 & 16 & 0.44 \\
        \hdashline
        13 & Security Guards & Administrative \& support services & Operate detecting devices to screen individuals and prevent passage of prohibited articles into restricted areas. & Touchless, automated and remote premise entry systems and methods & 0.8566 & 6 & 14 & 0.43 \\
        \hdashline
        14 & Remote Sensing Scientists and Technologists & Manufacturing & Develop automated routines to correct for the presence of image distorting artifacts, such as ground vegetation. & Method for plantation treatment based on image recognition & 0.8413 & 10 & 24 & 0.42 \\
        \hdashline
        15 & Machinists & Manufacturing & Machine parts to specifications, using machine tools, such as lathes, milling machines, shapers, or grinders. & Machining equipment system and manufacturing system & 0.8440 & 12 & 29 & 0.41 \\
        \hdashline
        16 & Radiologists & Healthcare & Perform or interpret the outcomes of diagnostic imaging procedures including magnetic resonance imaging (MRI), computer tomography (CT), positron emission tomography (PET), [...] & Systems and methods for integrating tomographic image reconstruction and radiomics using neural networks & 0.8569 & 7 & 17 & 0.41 \\
        \hdashline
        17 & Computer Numerically Controlled Tool Operators & Manufacturing & Implement changes to machine programs, and enter new specifications, using computers. & Registering collaborative configuration changes of a network element in a blockchain ledger & 0.8370 & 11 & 27 & 0.41 \\
        \hdashline
        18 & Atmospheric and Space Scientists & Scientific and technical services & Analyze historical climate information, such as precipitation or temperature records, to help predict future weather or climate trends. & Combining forecasts of varying spatial and temporal resolution & 0.8429 & 11 & 27 & 0.41 \\
        \hdashline
        19 & Textile Knitting and Weaving Machine Setters, Operators, and Tenders & Manufacturing & Set up, or set up and operate textile machines that perform textile processing [...] & Parameter Manager, Central Device and Method of Adapting Operational Parameters in a Textile Machine & 0.8477 & 8 & 20 & 0.40 \\
        \hdashline
        20 & Medical Transcriptionists & Healthcare & Transcribe dictation for a variety of medical reports [...] & Methods for improving natural language processing with enhanced automated screening for automated generation of a clinical summarization report and devices thereof & 0.8367 & 6 & 15 & 0.40 \\
        \hline
    \end{tabular}
    }
\end{table*}

\begin{figure*}[t!]
\begin{center}
    \centering
    \includegraphics[width=0.9\textwidth]{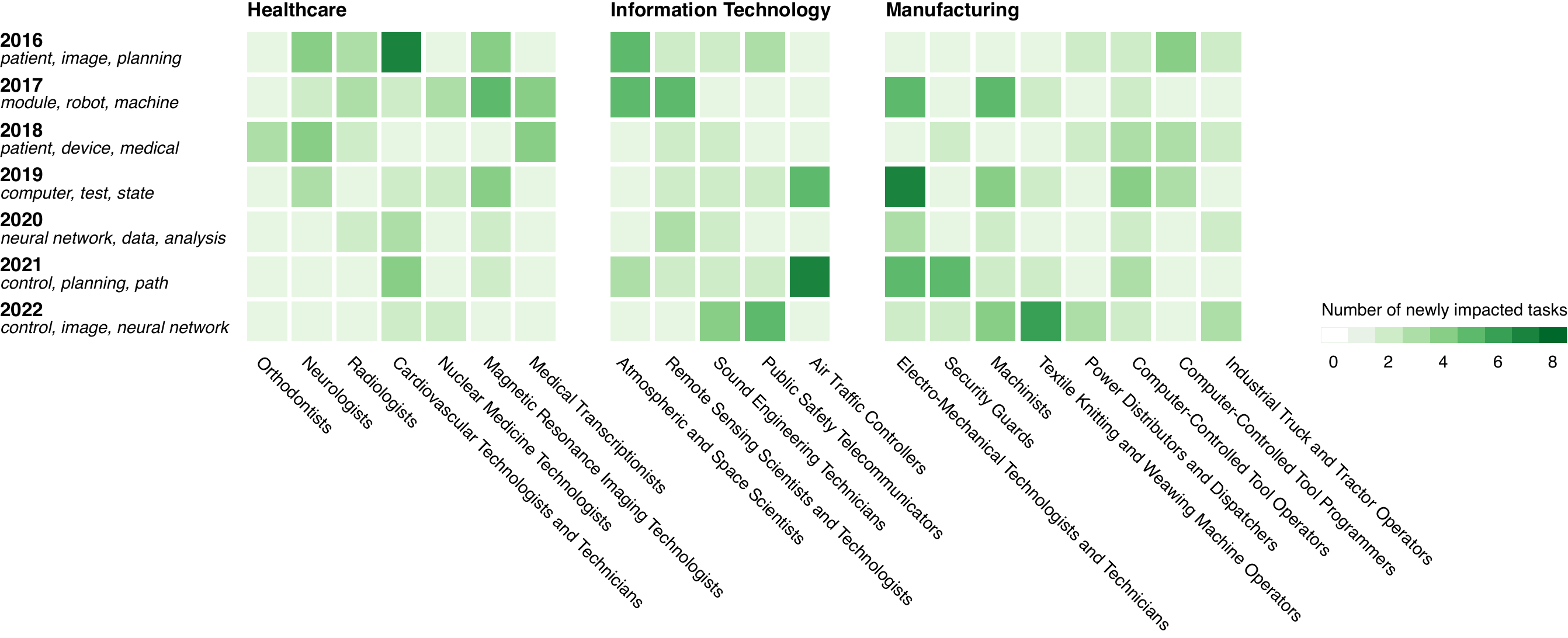}
    \caption{ The number of newly impacted tasks each year for the most affected occupations, combined with the most frequently occurring words in the patents influencing those tasks, are organized around the themes of healthcare, information technology, and manufacturing. These were derived qualitatively and describe the main themes emerging from the patents.
    Between 2016 and 2018, patents mentioned ``patient'', ``image'', ``planning'', ``medical'', ``device'' matched tasks in healthcare. Between 2019 and 2021, patents mentioned ``data'', ``analysis'', and ``neural networks'' matched tasks in information technology. Between 2021 and 2022, patents mentioned ``control'', ``planning'', ``path'', ``user'', ``image'', and ``neural network'' matched tasks in manufacturing.}
    \label{fig:occupations-temporal-impact-new}
\end{center}
\vspace{-0.2 in}
\end{figure*}

\subsubsection*{Most-impacted Occupations}
The highest-impact occupations mainly consist of white-collar occupations such as cardiovascular technicians, sound engineers, nuclear medicine technologists, air traffic controllers, and magnetic resonance imaging technologists. Indeed, the AII score, binned by education levels from high school to Master of Sciences, shows that the highest impact is seen for jobs requiring degrees from community colleges or Bachelor's degrees (Figure~\ref{fig:education}) rather than high school diplomas or lower qualifications. 
Highest-impacted occupations are primarily found in the healthcare, information technology, and manufacturing. Their tasks can be completed in a very specific sequence, and the inputs and outputs of these tasks can be expressed in a machine-readable format. To see how, we examined the types of tasks that patents have automated, and organized the patents into three themes: \emph{healthcare}, \emph{information technology}, and \emph{manufacturing} (Table~\ref{tab:job_tasks_patents}). For the theme of healthcare, from 2015 to 2022, 60\% of the tasks done by cardiovascular technologists and 48\% of those done by magnetic resonance imaging (MRI) technologists have been impacted by patents automating health records' management and analyzing MRI scans. In addition to the advanced healthcare occupations, we observed a significant number of patents impacting less skilled healthcare personnel including, for example, patents recording and evaluating patient questionnaires. For the theme of information technology, over the same five years, 47\% of tasks done by software developers and 40\% of those done by computer programmers have been impacted by patents automating programming tasks and developing workflows. For the theme of manufacturing, over the same five years, 45\% of tasks done by truck and tractor operators and 40\% of earth drillers' tasks have been automated. These automated tasks are planning processes such as water-well drilling rigs and driving through electric-powered trucks.


Studies on long-haul truck driving show that truckers are not being replaced by AI~\cite{levy2022robotruckers}. Instead, they are now using smart technology that monitors their health, like smartwatches and advanced health devices. Previous research did not account for these new technologies that came out over the years~\cite{felten2021occupational}. Our score, on the other hand, is adjusted based on new innovations, so it changes over time.

By analyzing the annual increase in impacted tasks and examining the most frequently occurring words in patents' abstracts each year (Figure~\ref{fig:occupations-temporal-impact-new}), we identified two categories of highly impacted occupations. The first category includes occupations that experienced a sudden impact, while the second category includes occupations that faced continuous impact over time. 

Occupations that underwent sudden impacts are predominantly within the healthcare sector. This  impact became most pronounced in 2016, when eight new tasks were affected, gradually decreasing to just one new impacted task by 2019. In 2016, patents began to significantly impact healthcare by automating medical imaging and diagnosis through machine learning models, devising treatment plans and medical devices, and recording and analyzing patient data. This impact continued into 2017 and 2018, although to a lesser degree, focusing on predicting optimal radiation therapy doses, dental treatment plans, and processing medical patient data.

Occupations that sustained continuous impact over time are primarily in information technology and manufacturing. In the information technology sector, occupations such as software developers, and in manufacturing, occupations such as earth drillers, saw consistent increases from zero new impacted tasks in 2016 to six new impacted tasks in 2019. In information technology, the potential impact of patents became noticeable in 2017 when they began training robots to design and execute iterative tests on chemical samples, working on aerial and satellite imagery to create products such as land cover maps, and implementing speech recognition and natural language processing on audio. This impact steadily rose and extended into 2022, with patents integrating machine learning into software systems, automating tasks such as troubleshooting networks and code reviews. In manufacturing, patent potential impact emerged in 2018, focusing on optimizing supply chain logistics and planning material dumping operations. This impact persisted into 2022, further supporting predictive maintenance and operational optimization such as determining aircraft conditions, with patents integrating reinforcement learning and other advanced neural networks.

\subsubsection*{Least-impacted Occupations}  
The least-impacted occupations mainly consist of blue-collar occupations such as pile-driver operators, dredge operators, aircraft cargo handling supervisors, agricultural graders and sorters, and insurance underwriters. Again, the AII score, binned by education levels, shows that the lowest impact is seen for jobs requiring a high school diploma or less (Figure~\ref{fig:education}). These occupations are primarily found in agriculture, transportation, accommodation and food services, and construction sectors, where the core tasks and responsibilities revolve around physical and manual labor and typically do not require a wide range of complex mental or physical activities, nor do they involve abstract reasoning. In agriculture, least-impacted tasks involve food inspection and dairy management for livestock. In transportation, such tasks involve scheduling and resource allocation for airline operations, engine sound control, and vehicle dispatching. In accommodation and food services, tasks include monitoring and recording food temperatures. In construction, such tasks involve the maintenance and operation of equipment and machinery. In addition to these sectors, by examining occupations with nearly zero AII scores, we found another least impacted set of occupations: managerial ones. Contrary to the previous least impacted occupations which involve manual labor or dexterity, managerial occupations typically require human interactions and scarce expert knowledge tacitly acquired over years of experience, having tasks ranging from contract negotiation to proposal review to internal assessments and audits.

\begin{figure}[t!]
    \centering
    \subfloat[]{{\includegraphics[width=.45\textwidth]{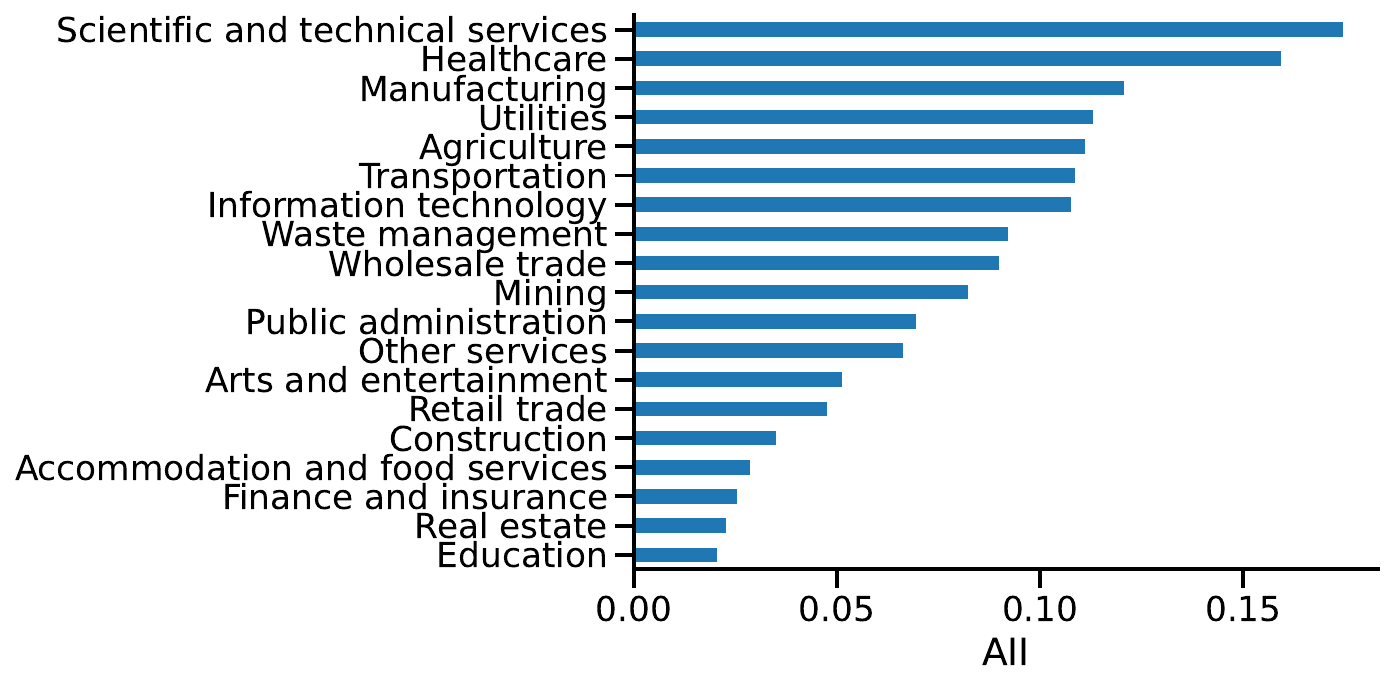} }}
    \qquad
    \subfloat[]{{\includegraphics[width=.45\textwidth]{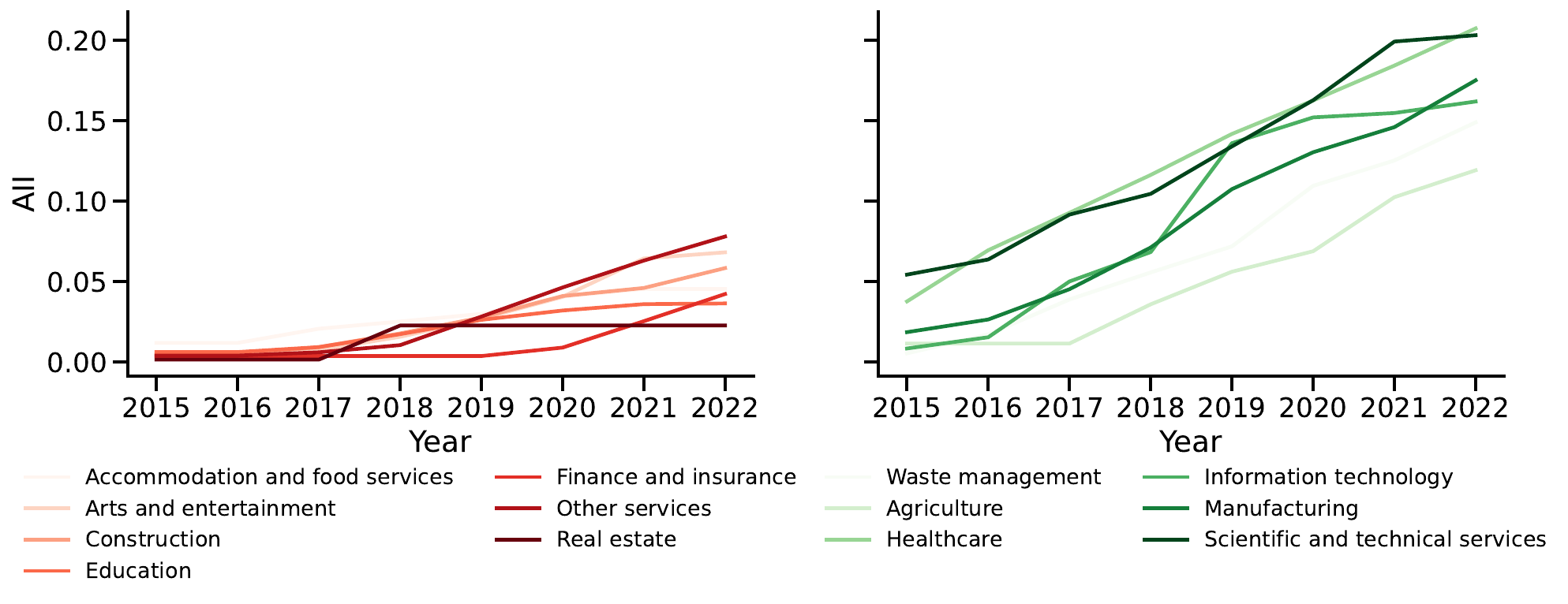} }}
    \caption{Sector-level AII scores for: (a) all sectors; (b, left) sectors with lowest rate of change from 2015 to 2020; and (b, right) sectors with highest rate of change.}
    \label{fig:industry_sectors}
\end{figure}

\begin{figure}[t!]
    \centering
    \includegraphics[width=.45\textwidth]{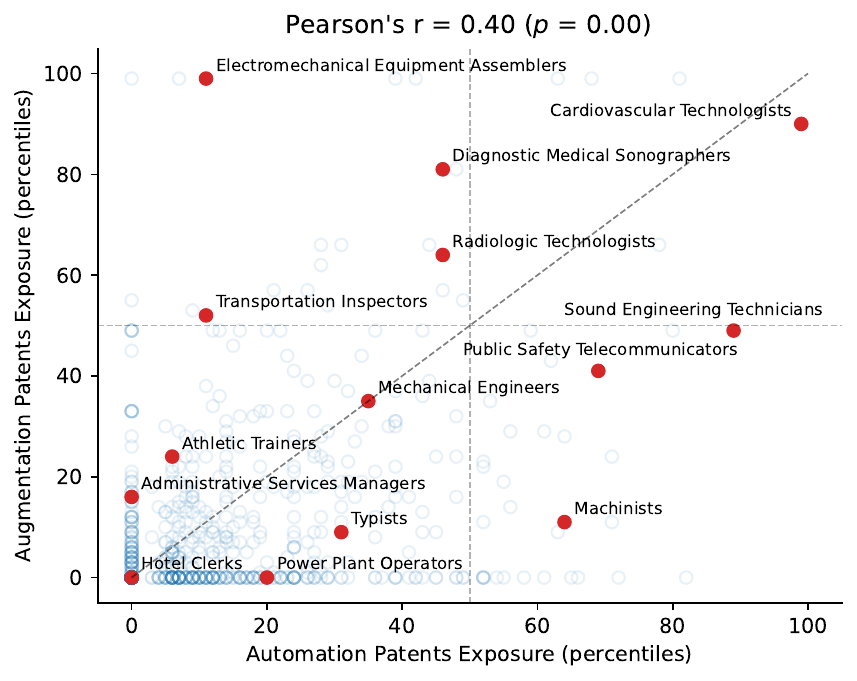}
    \caption{Automation \emph{vs.} augmentation using patent similarity to tasks and micro-titles defined in the Census Alphabetical Index of Occupations and Industries (CAI)~\cite{autor2022new}.}
    \label{fig:auto-aug-micro-titles}
\end{figure}

Overall, AI's impact on occupations is limited because: (1) some occupations are augmented by AI rather than replaced, and (2) the industries being affected already have a shortage of workers.

\subsection*{Augmented rather than replaced occupations}
\label{subsec:results-automation-augmentation}

Our AII captures AI's potential for automation. However, previous work has differentiated between automation and augmentation. To account for that, 
we implemented Autor \emph{et al.}'s method~\cite{autor2022new} to compute AI's potential for augmentation (explained in ``\nameref{sec:methodology}''). We found that certain occupations will not be replaced by AI, but instead will be augmented by it (top left quadrant in Figure~\ref{fig:auto-aug-micro-titles}). For example, the role of a hearing aid specialist involves a significant human element, especially in understanding and responding to patients' emotional, psychological, and physical needs. This is reflected in the lack of patents that match the task of ``counseling patients and families on communication strategies and the effects of hearing loss''; a task that requires empathy and emotional intelligence, which is likely no current AI innovation can automate. Another example is that of electrical and electronics repairers, which relies heavily on technical skills, detailed knowledge, and hands-on interaction. Consider that occupation's task of ``consulting with customers, supervisors, or engineers to plan the layout of equipment or to resolve problems in system operation or maintenance''. While automation can assist in some aspects, human expertise and decision-making are crucial.  This task has not been matched with any patent. In general, for occupations that may be augmented by AI, AI will advise, coach, and alert decision-makers as they apply expert judgment. 

\subsection*{Affected sectors experiencing labour shortages}
\label{subsec:impact_sectors}
\begin{figure}[t!]
    \centering
    \includegraphics[width=.4\textwidth]{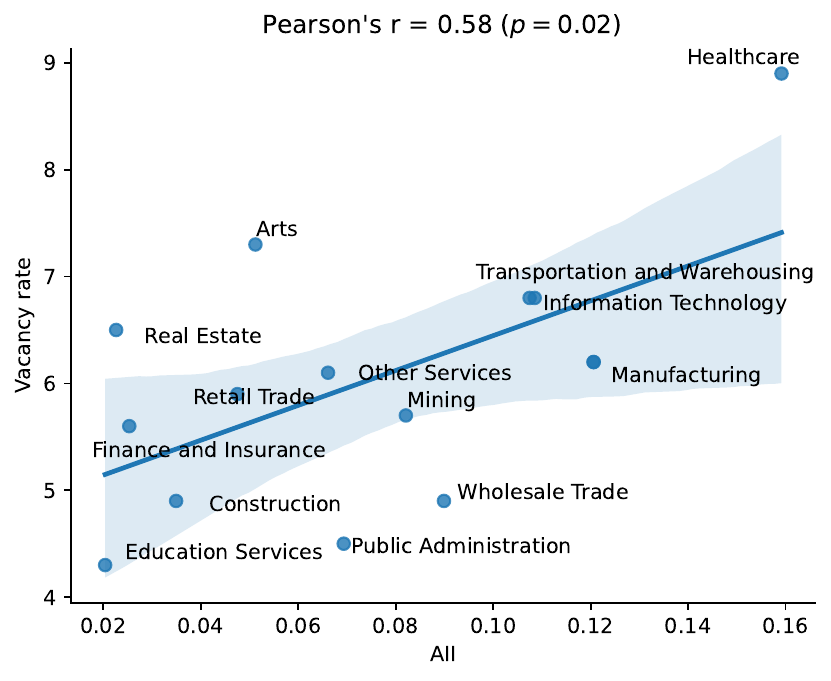} 
    \caption{Job vacancy rates by sector \emph{vs.} sector-level AII. The sector of Accommodation and Food Services was positioned more than two standard deviations away from the regression line (i.e., considered as an outlier) and was removed. The original plot is in Supplementary Material.}
    \label{fig:vacancy-sector}
\end{figure}

 To ascertain whether AI patents are linked to labor supply constraints or to labor demand in a sector, we correlated the AII industry sector scores with the annual vacancy rates from each sector in 2022. We found that AII and vacancy rates are positively correlated. After removing the outlier sector of Accommodation and Food Services---positioned more than two standard deviations away from the regression line (Figure~\ref{fig:vacancy-sector-outlier}) ---the correlation was positive and stronger, with a Pearson's correlation coefficient of  $r = 0.58$ with $p = 0.02$ (Figure~\ref{fig:vacancy-sector}). This suggests that sectors potentially impacted by AI are currently experiencing labor shortages.

Highly-impacted sectors include healthcare, information technology, and manufacturing, which aligns with our thematic analysis of tasks within the most-impacted occupations. These sectors have experienced a significantly high rate of impact. From the thematic analysis, two possible explanations emerge. The first explanation lies in the nature of the tasks within these sectors, which are likely to be replaced by AI-enhanced hardware. For example, in healthcare, tasks involving the use of X-rays or MRI scans, such as those used by radiologists, have been automated by patents on advanced medical equipment and devices. Similarly, in manufacturing, tasks that involve the examination of chemical or biological samples, such as those performed by food science technicians, can now be executed by by AI-enhanced hardware.  The second explanation is that the occupations within these sectors entail tasks demanding extensive data analysis and processing. In information technology, film editors, for example, engage in video data editing, which our patent analysis found to have been streamlined by AI-based software. Likewise, scientists in healthcare, information technology, or manufacturing often handle large volumes of data, and recent patents deal with both structured and unstructured data (e.g., using deep-learning for computational biology~\cite{angermueller2016deep}).

\section*{Discussion}
\label{sec:discussion}

\subsection*{Consensus and Discrepancies in the Literature}
Unlike previous methods for assessing the potential impact of AI on occupations, which either rely on a finite set of abilities (e.g., manual dexterity) linked to specific occupations~\cite{felten2021occupational, frey2017future} or employ subjective evaluations to determine tasks' suitability for automation~\cite{brynjolfsson2018machines}, our AII measure provides an objective approach by leveraging patent data to capture the dynamic landscape of technological advancements in AI. 

To position our findings, we explore the consensus or lack thereof of which occupations will be potentially impacted by AI in the literature.

\subsubsection*{Consistent results in the literature} Frey and Osborne's~\cite{frey2017future}, AIOE~\cite{felten2021occupational}, SML~\cite{brynjolfsson2018machines}, and Webb's~\cite{webb2019impact} concur that low-skilled occupations typically involving human labor are not impacted by AI. These occupations are typically found in industry sectors such as construction, where tasks primarily involve manual labor and the operation of machinery. Our AII measure aligns with these observations by categorizing such occupations as among the least impacted by AI. Least-impacted sectors include construction, accommodation and food services, real estate, education, public administration, and finance and insurance. These sectors have experienced a significantly low rate of impact. From the thematic analysis, three possible explanations emerge. The first explanation centers on the fact that occupations within certain sectors are often associated with low-skilled or physical/manual labor. For instance, in construction, tasks range from assembling solar panels to maintaining pipe systems to operating various drills, all of which require physical labor. Additionally, manual dexterity is challenging to automate. In public administration, numerous occupations, such as police officers and firefighters, still require manual skills. The second explanation is that occupations within certain sectors demand basic and non-specialized skills. For instance, in accommodation services, the tasks of waiters and baristas typically involve minimal specialized knowledge or vocational training, such as serving food and drinks. Similarly, in real estate, brokers or clerks primarily require training in overseeing transactions and handling tasks related to office operations.The third explanation is that occupations within certain sectors often require a high level of skill and involve extensive interpersonal interactions. For instance, in managerial positions (e.g., CEOs), tasks typically encompass responsibilities such as delegating tasks, attending events and meetings, and negotiating contracts - all of which heavily rely on human interpersonal communication. Similarly, in education, a teacher's role primarily revolves around delivering educational materials in person, a task that demands both physical presence and a higher level of education and specialized knowledge. In the legal sector, potentially impacted occupations include those involved in drafting legal documents or transcribing pretrial and trial proceedings, such as court reporters, in alignment with previous qualitative analyses on legal occupations~\cite{armour2020ai}. Conversely, roles requiring the design of bespoke legal solutions remain unaffected. Notably, client-facing tasks in the legal sector are also not impacted, resulting in an overall expectation of limited impact on the legal sector. Finally, in the financial sector, there are still occupations that necessitate human interactions, such as clerks, sales agents, and tellers.

\subsubsection*{Inconsistent results in the literature} 
While there has been a clear consensus in previous literature regarding manual labor occupations, we have identified four types of occupations for which consensus has been so far lacking. These categories include: occupations requiring basic and non-specialized skills; occupations requiring high skills and interpersonal interaction-based occupations; those where tasks are replaced by AI-enhanced hardware; and those involving extensive data analysis and processing.

\mbox{ }\\
\noindent
\emph{Occupations that require basic and non-specialized skills (least impacted according to our approach).} Frey and Osborne's findings~\cite{frey2017future} indicated that occupations within the accommodation and food services sector (e.g., cooks, dishwashers, waiters, bartenders) are highly-impacted. Similarly, Webb's findings suggest that non-routine manual occupations are highly impacted; there are patents matching those occupations, which may well be coarse-grained matches, as exemplified in Table~\ref{tab:elevator}. However, SML~\cite{brynjolfsson2018machines} and AIOE~\cite{felten2021occupational} found the very same occupations to be among the least-impacted. Our approach aligns with SML and AIOE, identifying these occupations as among the least-impacted due to the absence of AI patents automating the manual tasks associated with them.

\mbox{ }\\
\noindent
\emph{Occupations that are highly-skilled and involve interpersonal interactions (least impacted according to our approach).} AIOE~\cite{felten2021occupational} found that occupations requiring high skills and interpersonal interactions, such as those in the education sector (e.g., teachers) or managerial positions (e.g., CEOs), are highly-impacted. Similarly, Webb found that occupations that require interpersonal tasks are hard to automate~\cite{webb2019impact}. In contrast, SML~\cite{brynjolfsson2018machines} found these occupations, particularly managerial positions, to be among the least impacted, aligning with Frey and Osborne's method~\cite{frey2017future}. However, in the case of SML, there was a distinction within occupations involving interpersonal interactions. While SML found managerial positions to be among the least impacted, it identified occupations such as concierges, which also involve interpersonal interactions, as more likely to be impacted by AI. Upon closer examination of the SML method, which relies on rubrics (i.e., a type of scoring guide for crowdworkers to assess the suitability of tasks for machine learning, as shown in Table~\ref{tab:rubric}), we noted that the corresponding definition in the rubrics entailed a ``wide range of interactions,'' making it challenging to capture their nuances. Additionally, since SML relies on crowdsourced data, there is a potential for subjective bias or a lack of full understanding of the nuances of interpersonal interactions. Finally, interactions were captured by only one item out of the 23 items in the rubrics making that item's contribution to the overall score limited. Our approach aligns with Frey and Osborne's~\cite{frey2017future}, Webb's~\cite{webb2019impact} and, to some extent with SML~\cite{brynjolfsson2018machines} as no AI patents were found to target these occupations.

\mbox{ }\\
\noindent
\emph{Occupations that consist of tasks that are replaced by AI-enhanced hardware (most impacted according to our approach).} Frey and Osborne~\cite{frey2017future} discovered that occupations in healthcare, such as MRI and cardiovascular technologists, which are likely to be replaced by AI-enhanced hardware, were among the least-impacted. In contrast, SML~\cite{brynjolfsson2018machines} and AIOE~\cite{felten2021occupational} identified these same healthcare occupations as among the most-impacted ones. Webb~\cite{webb2019impact} found that occupations involving non-routine manual tasks (e.g., operating devices or equipment in healthcare or manufacturing) to be among the most-impacted ones. Our approach aligns with SML, AIOE, and Webb's method due to the presence of AI patents automating tasks using AI-enhanced hardware. For example, in healthcare, patents have automated tasks such as medical imaging and diagnosis using machine learning models, the development of treatment plans, the creation of medical devices, the recording and analysis of patient data, and even the prediction of optimal radiation therapy doses, dental treatment plans, and the processing of medical data. 

\mbox{ }\\
\noindent 
\emph{Occupations that require extensive data analysis and processing (most impacted according to our approach).} Frey and Osborne~\cite{frey2017future} found that occupations that require extensive data analysis and processing, such as those in information technology or manufacturing (e.g., software developers, chemists, aerospace engineers), were classified as among the least-impacted. In contrast, SML~\cite{brynjolfsson2018machines} and AIOE~\cite{felten2021occupational} categorized these very same occupations as among the most-impacted. Similarly, Webb~\cite{webb2019impact} identified that occupations involving non-routine cognitive analytic tasks to be among the most-impacted ones. Our findings align with SML, AIOE, and Webb's method due to the presence of patents related to tasks that require data analysis and processing. For example, patents about training robots for task execution, image and video processing, speech recognition, and natural language processing, as well as the integration of machine learning into software systems. These patents automate tasks such as troubleshooting networks, code reviews, optimizing supply chain logistics, planning material dumping operations, and supporting maintenance, including determining aircraft conditions. Our findings are further confirmed by recent works studying the potential exposure of occupations and tasks to Generative AI~\cite{gmyrek2023genai,eloundou2023gpts}. They found that Generative AI will potentially impact high-skilled, intellectual, or creative professions (e.g., mathematicians, writers, and translators), where these models can effectively augment capabilities in data analysis, writing, and language translation.

\subsection*{Implications}
The impact of AI on occupations carries important implications for the workforce. However, when placed our findings within the context of previous literature, it became evident that there is a lack of consensus regarding which occupations will be affected and which will remain unaffected. As an initial step, we contend that achieving consensus is crucial to start formulating effective policies to address the ongoing transformations in the labor market. To initiate this process, with our findings in mind, we outline three key areas in which initiatives can be developed.

\subsubsection*{Initiatives for White-Collar Workers} Policymakers and employers should launch specific initiatives targeting white-collar occupations in sectors such as information technology, manufacturing, and healthcare. These initiatives can equip workers with the skills needed for high-value, creative, and complex tasks. For example, a manufacturing worker can undergo training in robotics programming, enabling them to effectively operate and maintain AI-driven machinery. Similarly, healthcare professionals can acquire telemedicine and data analytics skills to enhance patient care and diagnostics.


\subsubsection*{Initiatives for Blue-Collar Workers} 
Blue-collar occupations, predominantly found in agriculture, accommodation and food services, and construction, typically involve low-skilled work demanding physical labor. While previous literature has suggested reallocating low-skilled workers to tasks less susceptible to computerization~\cite{frey2017future}, such as those requiring creative and social intelligence, we argue that (re)skilling and upskilling~\cite{jagannathan2019dominant} should be approached cautiously.

\subsubsection*{Initiatives for Continuous Learning and Interdisciplinary Training} 
Promoting a culture of continuous learning and skill development is essential. Employers can encourage individuals to embrace lifelong learning through online courses, certifications, and vocational training, enabling them to adapt to the changing occupational landscape~\cite{mckinsey_2016}. Similarly, encouraging interdisciplinary training can prepare the workforce for the demands of AI-augmented occupations. For example, blending traditional engineering with AI and machine learning training can create a workforce capable of developing and maintaining AI-enhanced systems across sectors.

\subsection*{Limitations and Future Work}
This work has five primary limitations. First, our analysis is conducted on an annual basis and assumes that the tasks associated with a given occupation in the O*NET database remain consistent across all versions within the same year.

Second, our method relies on patent abstracts to provide a finer-grained understanding of occupational tasks, mainly from the US-focused USPTO dataset. Despite the most significant innovations are typically patented in all major patent jurisdictions~\cite{mann2023benign}, the U.S. holds a distinct position. In 2014, nearly a quarter of the approximately 10.9 million patents worldwide were granted in the U.S., highlighting its significant share~\cite{wipo_2016}. However, more recently, Carbonero \emph{et al.}~\cite{carbonero2023developing} and Guarascio \emph{et al.}~\cite{guarascio2023artificial} studied the potential impact of AI on occupations in Southeast Asia and Europe. Unlike our method, their methods relied on manual annotations to determine
the suitability of specific tasks for machine learning. We therefore calculated the AII score based on US patents alone and compared it with the AII score from patents combined from the US, China, Japan, and South Korea, finding them to be highly correlated ($r = 0.93$), leaving our results unaffected. However, future research should still replicate our method and explore potential cultural differences in how patents are written and used in other contexts, not least in the European Union.  

Third, it is important to acknowledge that the existing portfolio of USPTO patents may not comprehensively cover all the innovations that may impact a particular occupation. To address this limitation, one potential approach could involve supplementing patent data with other sources, including research papers and code repositories~\cite{cao2023breaking}.

Fourth, our assumption is that the likelihood of an AI system being built is determined by whether it is patented. While generally true, there are exceptions. A patented system may not be built, as the patent could be intended for defensive or offensive purposes~\cite{blind2009influence, jell2017offensive}. Conversely, a non-patented system might still be constructed, with its design protected by secrecy or trademarks~\cite{cohen2000protecting, arundel2001relative}. While there are numerous patents for systems designed to improve meetings, calendaring, and instant messaging, patents focused on interpersonal interactions may be less common. However, even if these patents do not directly threaten jobs requiring human interaction, they could still have secondary effects. For example, if AI significantly affects occupations like artists or software developers, managerial positions may become less essential due to a reduced workforce needing oversight. Yet, our method does not capture such cascading impacts on the job market. Furthermore, patenting rates vary among sectors~\cite{patents_per_sector}, and there is a lag between an innovation being patented and its use and impacts diffusing across the economy~\cite{cao2023breaking}.

Finally, our analysis is based solely on a concise yet comprehensive seven-year time window, spanning from 2015 to 2022. We also repeated the analysis for 2010-2022 and found no significant difference in the results (as shown in Table~\ref{tab:job_aii_2022} and Figure~\ref{fig:sector-2022}). This does not capture emerging technologies, such as Large Language Models (LLMs). Future research could replicate our methodology to assess the potential impact of emerging technologies, such as cryptocurrency, the metaverse, and LLMs by using upcoming patents in those fields.

\section*{Materials and Methods}
\setcurrentname{Materials and Methods}\phantomsection\label{sec:methodology}

\subsection*{Datasets}
\setcurrentname{Datasets}\phantomsection\label{subsec:datasets}

\subsubsection*{Occupation Dataset} We collected detailed task descriptions for a wide range of occupations from the O*NET database~\cite{onet2023database}, a widely used source in occupational studies~\cite{goos2009job, autor2013tasks, brynjolfsson2018machines, felten2021occupational, felten2023chatgpt}. In total, we collected 759 unique occupations and 17,879 unique tasks from O*NET 26.3 version released in May 2022. The distribution of these tasks ranges from 4 to 286, with a median of 20 tasks per occupation (Figure~\ref{fig:task-histogram}).

\subsubsection*{Patent Dataset} To obtain a corpus of AI patents, we first retrieved 74,875 patents granted by the United States Patent and Trademark Office (USPTO) between 2015-2022 that were classified to be about AI based on an official way of coding patents called PATENTSCOPE AI Index~\cite{wipo2019ai} to then filter away patents only tangentially related to AI. We selected the subset of patents in the index class core AI applications (Table~\ref{tab:keywords}). This resulted in a final corpus of 24,758 AI patents. 

\subsection*{Measuring AI Impact (AII) on Occupation Tasks}
\setcurrentname{Measuring AI Impact (AII) on Occupation Tasks}\phantomsection\label{subsec:new_method}

\subsubsection*{Sentence-Transformers}
We developed a Sentence-Transformers Deep Learning framework \cite{reimers2019sbert} for Natural Language Processing that uses the Sentence-T5 (ST5) architecture \cite{ni2021st5} to convert input text into ``semantic vector representations'' called ``embeddings''. These embeddings capture the semantic information of the text and allow us to mathematically compute the similarity of a pair of text snippets. In particular, we chose the Sentence-T5-XL model, which is highly recommended for its effectiveness in handling various language tasks such as classification and similarity comparisons~\cite{reimers2019sbert}. This model has demonstrated exceptional performance in a comprehensive benchmark test---the Massive Text Embedding Benchmark---that evaluated different models across 58 datasets and 112 languages for embedding tasks such as classification, clustering, and semantic textual similarity~\cite{muennighoff2022mteb}. We used the default parameters of the model because they were already optimized for textual similarity tasks similar to ours. The model's default training parameters include an Adafactor optimizer at a learning rate starting at 0.0001, with linear decay after 10\% of the total training steps; the fine-tuning was conducted using a batch size of 2048, and a softmax temperature $\tau$ of 0.01 was used. The model was trained on a  dataset of 2 billion question and answer (Q\&A) pairs from online Q\&A communities, and was then fine-tuned to enhance its understanding of how sentences are related to each other by training on pairs of sentences that had been manually reviewed for their meaning~\cite{bowman-etal-2015-large}. The model uses a siamese network architecture~\cite{schroff2015facenet}, which processes pairs of sentences to generate a consistent output length, regardless of the sentence length. This method of producing fixed-size feature vector representations has proven effective in capturing the deeper meanings of text without the need for any preprocessing~\cite{le2014distributed,alzahrani2021different}.

\subsubsection*{AI Impact (AII)} We defined AII as a measure of \emph{``the exposure to AI by measuring the extent to which an occupation's tasks are associated with patents''}. For each task, we identified the patent with the highest task-patent similarity score (Equation~\ref{eq:aii}) to represent the AI potential impact $\alpha_t$ on task $t$ indicating the extent to which task $t$ aligns with AI-related innovations:
\begin{equation}
    \label{eq:aii}
    \alpha_t = \max_p \text{sim}(v_t, v_p)
\end{equation}
where $\text{sim}(v_t, v_p)$ is the cosine similarity between the embeddings of task $t$ and the embeddings of patent $p$. We computed the impact of AI on task $t$ by taking the maximum similarity value. We took the maximum instead of, say, the average because if the average was used, similarity scores from patents that are not particularly relevant to the task would be factored into the calculation, thereby diluting the AI impact score (Figure~\ref{fig:similarity-dist}). Multi-instance learning was also considered as an alternative, but it did not produce any more accurate task-patent matching (explained in ``\nameref{appendix:validation}'' in the Supplementary Material).

\subsection*{Measures of AII on Occupations and Industry Sectors}
\setcurrentname{Measures of AII on Occupations and Industry Sectors}\phantomsection\label{subsec:measures}

\subsubsection*{AI Impact on Occupations.} We computed the AI impact $x_j$ on occupation $j$ by computing the number of \emph{AI-impacted tasks} over the total number of $j$'s tasks: 
\begin{equation}
    \label{eq:weighted_job}
    x_j = \frac{\sum_{t \in \text{tasks}(j)} 1_{\alpha_{t} > p_{90}(\alpha)}}{\sum_{t \in \text{tasks}(j)} 1}
\end{equation}
where $p_{90}(\alpha)$ is the 90\textsuperscript{th} percentile of AI impact values computed on all occupations' tasks, and $1_{\alpha_t > p_{90}(\alpha)}$ is an indicator function whose value is 1, if $\alpha_t > p_{90}(\alpha)$, and 0 otherwise. In other words, the AII measure is based on counting an occupation's tasks impacted by  AI 
without accounting  for the relative importance of each task, in a way similar to previous work~\cite{autor2022new}. Using the 90\textsuperscript{th} percentile as the threshold makes the AI impact measure more robust to noise, which was also suggested in a previous study \cite{chaturvedi2023automation}. Given that every task is assigned a similarity value in the previous step, the patent deemed most similar for a specific task might still be unrelated to that task. On the other hand, a higher 95\textsuperscript{th} percentile threshold would be too strict, as 55\% of the occupations would have zero impacted tasks. To further validate our task-patent matching method, two authors independently assessed the relevance of a patent to a task in a random sample of 100 task-patent pairs. Overall, their agreement was nearly perfect, with a Cohen's Kappa of 0.84.

\subsubsection*{AI Impact on Industry Sectors}
To determine the potential impact of AI on industry sector $s$, we calculated the mean AII score across all occupations $j$ associated with sector $s$ (Equation~\ref{eq:proportion_industry}). Occupation $j$ was assigned to sector $s$, if over 50\% of workers in $j$ were employed in $s$:
\begin{equation}
    \label{eq:proportion_industry}
    \pi_s = \frac{1}{N_s}\sum_{j \in \text{occupations}(s)} x_j
\end{equation}
where $N_s$ is the number of occupations associated with $s$. If more than half of the workers in an occupation are employed in a particular sector, it can be reasonably concluded that this occupation is primarily associated with that sector \cite{felten2021occupational}. Lower thresholds might lead to occupations being associated with multiple sectors, making the results less specific. Conversely, a higher threshold might be too restrictive, potentially excluding occupations with a significant presence in a sector, even if not overwhelmingly so.

\subsection*{Thematic Analysis on Occupations and Industry Sectors}
\setcurrentname{Thematic Analysis on Occupations and Industry Sectors}\phantomsection\label{subsec:thematic}

\subsubsection*{Occupations} To identify emergent themes that characterize least or most impacted occupations, we conducted a thematic analysis~\cite{ saldana2015coding, miles1994qualitative} on the task-patent pairs for all tasks associated with those occupations. This process consists of two steps: open coding, in which textual data is broken up into discrete parts and then coded; and axial coding, in which the researcher draws connections between the generated codes~\cite{braun2006using}. We first applied open coding to identify key concepts that emerged across multiple task-patent pairs; specifically two of the authors read all task-patent pairs, and marked them with keywords that reflected the key concepts expressed in the text. They then used axial coding to identify relationships between the most frequent keywords to summarize them in semantically cohesive themes (e.g., healthcare, information technology, and manufacturing). Themes were reviewed in a recursive manner rather than linear, by re-evaluating and adjusting them as new task-patent pairs were parsed.

\subsubsection*{Industry Sectors} Just as with occupations, we conducted thematic analysis~\cite{ saldana2015coding, miles1994qualitative,braun2006using} to uncover potential explanations for the most- and least-impacted sectors. We again applied open coding to identify key concepts that emerged on the descriptions of occupations (examples presented in Table~\ref{tab:jobs}) across the industry sectors under study, with the same two authors reading all descriptions and marking them with keywords that reflected the key concepts expressed in the text. We then again used axial coding to identify relationships and potential explanations.

\subsection*{Beyond Automation: Measuring Augmentation}
\label{subsec:automation-augmentation}

Automation refers to ``technologies that substitute for the labor inputs of occupations, potentially replacing workers performing these tasks''~\cite{autor2022new}.  That is what our AII captures, and it does so by first identifying the patent most similar to an occupation task (Equation~\ref{eq:aii}), and then computing the number of AI-impacted tasks over the total number of tasks at a given occupation (Equation~\ref{eq:weighted_job}). In addition to automation, previous research introduced a complementary type of AI's potential impact: augmentation. This refers to ``technologies that increase the capabilities, quality, variety, or utility of the outputs of occupations, potentially generating new demands for worker expertise and specialization''~\cite{autor2022new}. To measure augmentation, instead of measuring the similarity between patents and occupation tasks, we measured the similarity between patents and micro-titles defined in the Census Alphabetical Index of Occupations and Industries (CAI)~\cite{autor2022new},  and then computed the number of AI-impacted micro-titles over the total number of micro-titles at a given occupation. Unlike  occupation tasks, micro-titles   capture the ``emergence of new work categories that typically reflect the development of novel expertise within existing work activities (e.g., electrical trades skills specific to solar installations) or an increase in the market scale of a niche activity (e.g., nail care)~\cite{autor2022new}.'' For example, USPTO patent US20180275314A1 for ``method and system for solar power forecasting'' was linked to the micro-title of ``solar thermal installer'' and the task of ``performing computer simulation of solar photovoltaic (PV) generation system performance or energy production to optimize efficiency''. Similarly, patent US2022083792A1 for a ``method and device for providing data for creating a digital map'' was linked to the micro-title ``digital cartographer'' and the task of ``mapping forest tract data using digital mapping systems''. In addition to measuring automation (using our AII measure on occupation tasks) and augmentation (adapting our AII measure on micro-titles), we replicated the method proposed by Gmyrek \emph{et al.}~\cite{gmyrek2023genai}. This method uses the mean and standard deviation of task-level scores to distinguish between automation and augmentation (explained in ``\nameref{appendix:automation-augmentation}'' in Supplementary Material).

\section*{Acknowledgments}
We thank the anonymous reviewers for their valuable suggestions. We also thank the Social and Responsible AI team at Nokia Bell Labs Cambridge, David Autor, Mari Sako, Daron Acemoglu, Diane Coyle for their useful feedback on the manuscript. 

\section*{Funding}
Nokia Bell Labs provided support in the form of salaries for authors [AAS, MC, DQ], but did not have any additional role in the study design, data collection and analysis, decision to publish, or preparation of the manuscript. The specific roles of these authors are articulated in the `author contributions' section. 

\section*{Author contributions statement}
A.A.S., M.C., and  D.Q. designed research; A.A.S. performed research; A.A.S. and M.C. analyzed data; and A.A.S., M.C., and D.Q. wrote the paper.

\section*{Data Availability}
The occupation data supporting this study's findings are available from O*NET~\cite{onet2023database}, at \url{https://www.onetcenter.org/db_releases.html}. The patent data are available in Google Patents Public Data~\cite{google_patents} at \url{https://console.cloud.google.com/marketplace/product/google_patents_public_datasets/google-patents-public-data}, and can be retrieved using the query available in the project's page. The Quarterly Census of Employment and Wages (QCEW) data are made available by the U.S. Bureau of Labor Statistics at \url{https://www.bls.gov/cew/downloadable-data-files.htm}. All the datasets are compiled on the project's page at \url{https://social-dynamics.net/aii}.

\section*{Code Availability}

Code necessary to reproduce the analyses in this study is available in the project's page at \url{https://social-dynamics.net/aii}

\bibliographystyle{plain}
\bibliography{main}

\clearpage

\section*{Supplementary Material}
 
\setcounter{figure}{0} 
\renewcommand{\thefigure}{S\arabic{figure}} 
\setcounter{table}{0} 
\renewcommand{\thetable}{S\arabic{table}} 

\section*{Data}
\label{appendix:data}

\subsection*{Data Collection}

In a total of 24,758 AI patents granted between 2015 and 2022, the majority contained the keyword machine learning (46\%), followed by the keyword neural network (32\%), artificial intelligence (9\%), and deep learning (6\%) (Table~\ref{tab:keywords}).

\begin{table}[h!]
    \centering
    \caption{Number of patents based on keywords.}
    \label{tab:keywords}
    \begin{tabular}{|l|c|}
    \hline
    Keywords & Number of patents \\
    \hline
    machine learning & 10904 \\
    \hdashline
    neural network & 9364 \\
    \hdashline
    artificial intelligence & 2674 \\
    \hdashline
    deep learning & 1848 \\
    \hdashline
    planning & 1050 \\
    \hdashline
    natural language processing & 917 \\
    \hdashline
    reinforcement learning & 506 \\
    \hdashline
    computer vision & 463 \\
    \hdashline
    speech processing & 126 \\
    \hdashline
    predictive analytics & 69 \\
    \hdashline
    robotics & 64 \\
    \hdashline
    control methods & 29 \\
    \hdashline
    knowledge representation & 24 \\
    \hline
    \end{tabular}
\end{table}

\begin{figure}[h!]
    \centering
    \includegraphics[width=1.\linewidth]{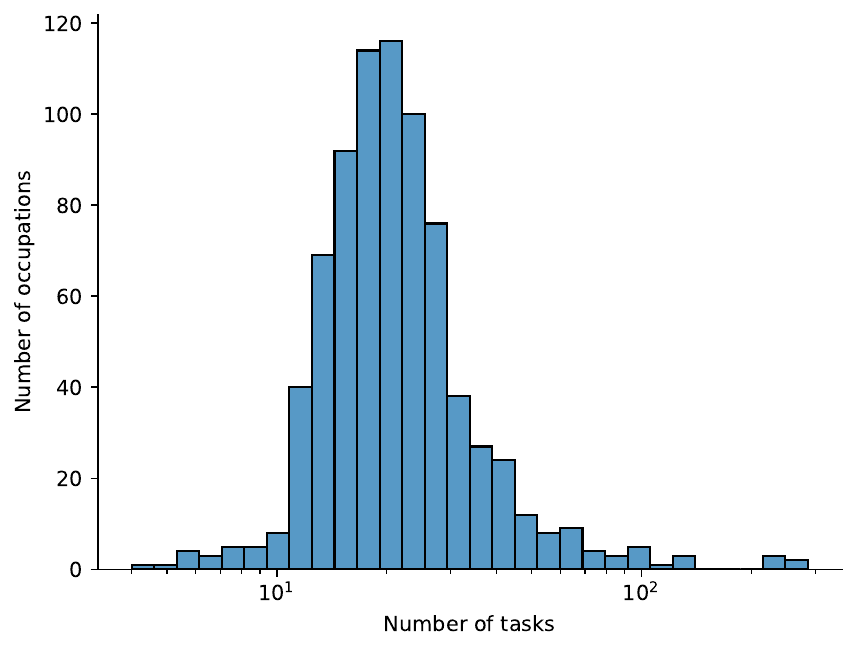}
    \caption{The distribution of the number of tasks per occupation.}
    \label{fig:task-histogram}
\end{figure}

From 759 occupations defined in O*NET, the number of tasks for each occupation ranges from 4 to 286, with a median of 20 tasks (Figure~\ref{fig:task-histogram}).

\clearpage

To compute SML scores, Brynjolfsson et al.~\cite{brynjolfsson2018machines} used a set of rubrics to guide crowdworkers to assess the suitability of tasks for machine learning (Table~\ref{tab:rubric}). Note that only one item in the rubrics is about interpersonal interactions.

\begin{table}[h!]
    \centering
    \small
    \caption{Rubrics to compute SML scores as developed by Brynjolfsson et al.~\cite{brynjolfsson2018machines}.}
    \label{tab:rubric}
    \scalebox{0.9}{
    \begin{tabular}{|c|p{17cm}|}
    \hline
    No & Rubric \\
    \hline
    Q1 & Information needed to complete the task (inputs) and outputs can be explicitly specified in machine-readable format \\
    \hdashline
    Q2 & Task information is recorded or recordable by computer \\
    \hdashline
    Q3 & Task does not require a wide range of complex outputs (mental and/or physical) \\
    \hdashline
    Q4 & Task feedback (on the success of outputs) is immediate \\
    \hdashline
    Q5 & The task output is error tolerant \\
    \hdashline
    Q6 & It is not important that outputs are perceived to come from a human \\
    \hdashline
    Q7 & Task does not require complex, abstract reasoning \\
    \hdashline
    Q8 & Task is principally concerned with matching information to concepts, labels, predictions, or actions \\
    \hdashline
    Q9 & Task does not require detailed, wide-ranging conversational interaction with a customer or other person \\
    \hdashline
    Q10 & Task is highly routine and repeated frequently \\
    \hdashline
    Q11 & Task is describable with rules \\
    \hdashline
    Q12 & There is no need to explain decisions during task execution \\
    \hdashline
    Q13 & Task can be converted to answering multiple choice questions, ranking alternative options, predicting a number, or grouping similar objects \\
    \hdashline
    Q14 & Long term planning is not required to successfully complete the task \\
    \hdashline
    Q15 & The task requires working with text data or might require working with text in the future \\
    \hdashline
    Q16 & The task requires working with image/video data or might require working with image/video data in the future \\
    \hdashline
    Q17 & The task requires working with speech data or might require working with speech data in the future \\
    \hdashline
    Q18 & The task requires working with other types of data (other than text, image/video, and speech) \\
    \hdashline
    Q19 & Many components of the task can be completed in a second or less \\
    \hdashline
    Q20 & Each instance, completion, or execution of the task is similar to the other instances in how it is done and these actions can be measured \\
    \hdashline
    Q21 & Actions in the task must be completed in a very specific order, and practicing the task to get better is easy \\
    \hline
    \end{tabular}
    }
\end{table}

\clearpage

To conduct the thematic analysis on industry sectors, we used the occupation titles and descriptions in O*NET. For brevity, we provide 10 examples in Table~\ref{tab:jobs} and we refer the reader to the full list in O*NET~\cite{onet2023database}.

\begin{table*}[tbp]
    \centering
    \caption{Examples of occupational descriptions in O*NET.}
    \label{tab:jobs}
    \scalebox{0.8}{
        \begin{tabular}{|p{4cm}|p{2.2cm}|p{11cm}|}
        \hline
        Occupation Title & Sector & Description \\
        \hline
        Cardiovascular Technologists and Technicians & Healthcare & Conduct tests on pulmonary or cardiovascular systems of patients for diagnostic purposes. May conduct or assist in electrocardiograms, cardiac catheterizations, pulmonary functions, lung capacity, and similar tests. Includes vascular technologists. \\
        \hdashline
        Orthodontists & Healthcare & Examine, diagnose, and treat dental malocclusions and oral cavity anomalies. Design and fabricate appliances to realign teeth and jaws to produce and maintain normal function and to improve appearance. \\
        \hdashline
        Medical Records and Health Information Technicians & Healthcare & Compile, process, and maintain medical records of hospital and clinic patients in a manner consistent with medical, administrative, ethical, legal, and regulatory requirements of the health care system. Process, maintain, compile, and report patient information for health requirements and standards in a manner consistent with the healthcare industry's numerical coding system. \\
        \hdashline
        Numerical Tool and Process Control Programmers & Manufacturing & Develop programs to control machining or processing of metal or plastic parts by automatic machine tools, equipment, or systems. \\
        \hdashline
        Multi-Media Artists and Animators & Information & Create special effects, animation, or other visual images using film, video, computers, or other electronic tools and media for use in products or creations, such as computer games, movies, music videos, and commercials. \\
        \hdashline
        Magnetic Resonance Imaging Technologists & Healthcare & Operate Magnetic Resonance Imaging (MRI) scanners. Monitor patient safety and comfort, and view images of area being scanned to ensure quality of pictures. May administer gadolinium contrast dosage intravenously. May interview patient, explain MRI procedures, and position patient on examining table. May enter into the computer data such as patient history, anatomical area to be scanned, orientation specified, and position of entry. \\
        \hdashline
        Nuclear Medicine Technologists & Healthcare & Prepare, administer, and measure radioactive isotopes in therapeutic, diagnostic, and tracer studies using a variety of radioisotope equipment. Prepare stock solutions of radioactive materials and calculate doses to be administered by radiologists. Subject patients to radiation. Execute blood volume, red cell survival, and fat absorption studies following standard laboratory techniques. \\
        \hdashline
        Software Developers, Applications & Information & Develop, create, and modify general computer applications software or specialized utility programs. Analyze user needs and develop software solutions. Design software or customize software for client use with the aim of optimizing operational efficiency. May analyze and design databases within an application area, working individually or coordinating database development as part of a team. May supervise computer programmers. \\
        \hdashline
        Electro-Mechanical Technicians & Manufacturing & Operate, test, maintain, or calibrate unmanned, automated, servo-mechanical, or electromechanical equipment. May operate unmanned submarines, aircraft, or other equipment at worksites, such as oil rigs, deep ocean exploration, or hazardous waste removal. May assist engineers in testing and designing robotics equipment. \\
        \hdashline
        Industrial Truck and Tractor Operators & Manufacturing & Operate industrial trucks or tractors equipped to move materials around a warehouse, storage yard, factory, construction site, or similar location. \\
        \hline
        \end{tabular}
    }
\end{table*}

\clearpage
\section*{Task-Patent Matching}
\setcurrentname{Task-Patent Matching}\phantomsection\label{appendix:validation}

\subsection*{Validating the similarity metric used for matching}

\begin{figure}[h!]
    \centering
    \includegraphics[width=\linewidth]{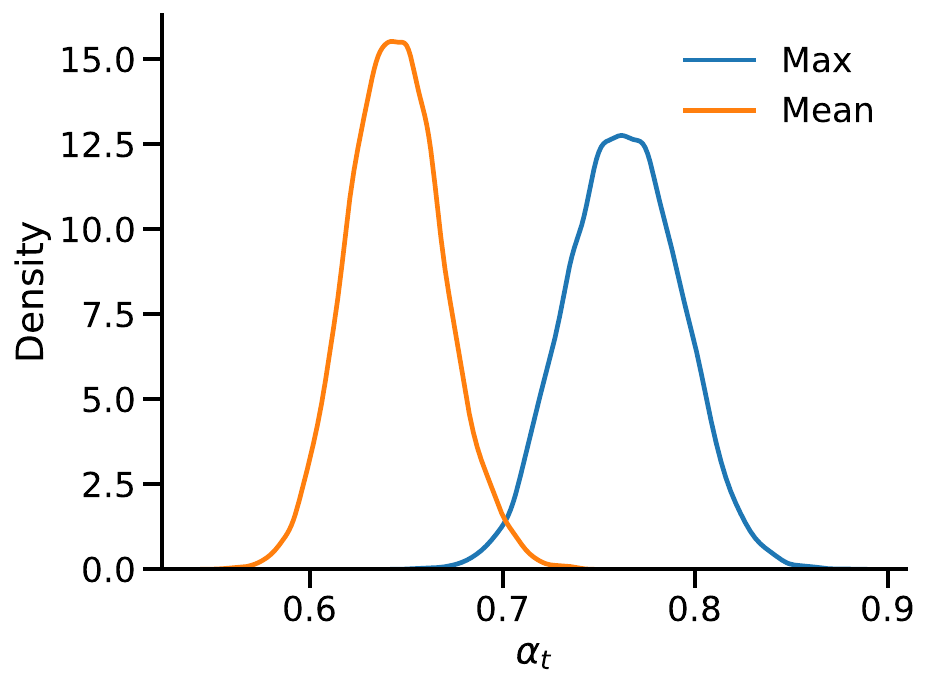}
    \caption{\textbf{Distributions of the task-patent similarity based on different aggregations.} By focusing only on the most relevant patents, we captured the specificity of the AI impact measurement for each task, reducing noise, and incorporating a wider range of AI-related advancements.}
    \label{fig:similarity-dist}
\end{figure}

To ensure the specificity of our task-patent matching method, we plotted the mean and max of the AII measure. We empirically found that by using the max, we retained the most relevant patent (Figure~\ref{fig:similarity-dist}). To further validate our task-patent matching method, we manually annotated a random sample of 100 task-patent pairs in terms of task-patent relevance; two of the authors who did the annotations achieved a Cohen's kappa of 0.84 (Kappa values in the [0.81, 1] range suggest almost perfect agreement).

\subsection*{Task-patent matching using multi-instance learning}
\label{appendix:multi-instance}

As an alternative to our AII measure for task-patent matching, we implemented a multi-instance learning approach. To do so, we followed a seven-step procedure:
\smallskip

\noindent\textbf{Step 1. Selection of Data:} We selected 100 patent abstracts, each related to just one specific task, and ensured that every abstract was closely linked to the task it described.
\smallskip
\noindent\textbf{Step 2. Sentence Segmentation:} We chunked these abstracts into individual sentences, resulting in a total of 356 sentences. 
\smallskip

\noindent \textbf{Step 3. Embedding Generation for Sentences:} We then generated embeddings for each of the individual sentences.
\smallskip

\noindent\textbf{Step 4. Embedding Generation for Tasks:} Simultaneously, we generated embeddings for the tasks associated with patent abstracts selected in Step 1. This parallel embedding process ensured that both the tasks and the patent sentences were represented in a comparable semantic space.
\smallskip

\noindent\textbf{Step 5. Pairwise Similarity Calculation:} With these embeddings at hand, we then computed the pairwise similarity between each task and all sentences from the abstracts related to it. This step involved calculating the cosine similarity between embeddings, which quantifies how semantically close two pieces of text are.
\smallskip

\noindent\textbf{Step 6. Maximum of Similarity Values:} For each task, we selected the maximum similarity value from the set of pairwise similarities. This value represents the closest semantic match between the task and sentences in the patent abstracts.
\smallskip

\noindent\textbf{Step 7. Comparison of Similarities:} Finally, we compared these maximum sentence-level similarity values with the similarity values obtained from whole abstract embeddings. This comparison allowed us to determine whether analyzing sentences individually provides a better match to the tasks than analyzing whole abstracts.

Following this seven-step procedure, we found a high correlation between similarity scores computed using embeddings of the entire abstract and those using embeddings of individual abstract sentences ($\rho = 0.85$). Moreover, 99 of the 100 selected tasks showed a higher similarity score with whole-abstract embeddings compared to the best-matching sentence. This suggests that using the entire abstract provides more context and leads to better matches than using the multi-instance learning approach.
    
\clearpage
\subsection*{Matching abstracts (not only titles)}
Patents tend to be general, and specific tasks may not be mentioned simply in the patent title. To address this, our method uses the patent abstract that is richer in text and more likely to reference a specific task or even entire sector(s) (Table~\ref{tab:generic-patents}). As a result, we were able to match the same patent with occupations in multiple sectors.
For example, a patent that provides methods for radiotherapy treatment planning not only matches the task of radiation therapists, but also counselors (abuse and  behavioral disorder) who need treatment plans. Another example is a patent for custom-fitting a service solution to consumer requirements that is applicable to occupations from different sectors (e.g., software developers, social workers, and solar photovoltaic installers).
\begin{table*}[htbp]
    \centering
    \caption{Example task-patent pairs matched with AII. The dash (-) sign means that the patent is the same as the cell above it.}
    \label{tab:generic-patents}
    \scalebox{0.6}{
    \begin{tabular}{|p{4cm}|p{2.2cm}|p{4cm}|p{4cm}|p{7cm}|c|}
        \hline
        Occupation & Sector & Task & Matching Patent Title & Matching Patent Abstract & Similarity \\
        \hline
        Substance Abuse and Behavioral Disorder Counselors & Healthcare & Develop client treatment plans based on research, clinical experience, and client histories. & Methods and systems for radiotherapy treatment planning & Example methods for radiotherapy treatment planning are provided. One example method may include obtaining training data that includes multiple treatment plans associated with respective multiple past patients; [...] & 0.8018 \\
        \hdashline
        Mental Health Counselors & Healthcare & Develop and implement treatment plans based on clinical experience and knowledge. & - & - & 0.8082 \\
        \hdashline
        Radiation Therapists & Healthcare & Administer prescribed doses of radiation to specific body parts, using radiation therapy equipment according to established practices and standards. & - & - & 0.8330 \\
        \hline
        Architectural and Civil Drafters & Manufacturing & Reproduce drawings on copy machines or trace copies of plans and drawings, using transparent paper or cloth, ink, pencil, and standard drafting instruments. & Line drawing generation & Computing systems and computer-implemented methods can be used for automatically generating a digital line drawing of the contents of a photograph. [...] The training data set teaches the neural network to trace the edges and features of objects in the photographs, as well as which edges or features can be ignored. [...] & 0.8038 \\
        \hdashline
        Multi-Media Artists and Animators & Information & Create pen-and-paper images to be scanned, edited, colored, textured, or animated by computer. & - & - & 0.8396 \\
        \hdashline
        Detectives and Criminal Investigators & Public administration & Create sketches and diagrams by hand or with computer software to depict crime scenes. & - & - & 0.8195 \\
        \hdashline
        Photographic Process Workers and Processing Machine Operators & Other services & Operate scanners or related computer equipment to digitize negatives, photographic prints, or other images. & - & - & 0.8167 \\
        \hline
        Software Developers, Applications & Information & Consult with customers about software system design and maintenance. & System and method for custom-fitting services to consumer requirements & Systems and methods for custom-fitting a service solution to consumer requirements are provided. [...] & 0.8020 \\
        \hdashline
        Child, Family, and School Social Workers & Education & Interview clients individually, in families, or in groups, assessing their situations, capabilities, and problems to determine what services are required to meet their needs. & - & - & 0.8010 \\
        \hdashline
        Solar Photovoltaic Installers & Construction & Determine photovoltaic (PV) system designs or configurations based on factors such as customer needs, expectations, and site conditions. & - & - & 0.8031 \\
        \hline
    \end{tabular}
    }
\end{table*}

\clearpage

\clearpage
\section*{Previous Attempts to Link Tasks to Patents}
\setcurrentname{Previous Attempts to Link Tasks to Patents}\phantomsection\label{appendix:selected-occupations}

\noindent\textbf{Word-matching methods linking tasks to patents.}
A word-matching method employs a dictionary approach, parsing task descriptions to identify verb-noun pairs associated with each task~\cite{webb2019impact}. This approach captures the task essence concisely and specifically, such as the pair ``install, sensor'', offering a clear representation of the task. Using the same method, AI patent titles are also processed to extract verb-noun pairs describing the tasks targeted by each patent. The relative frequency of similar pairs in tasks and patent titles determines the AI exposure score, indicating the level of task exposure to AI.

\mbox{ } \\
\noindent\textbf{Word-matching methods: Potential false positives.} Webb method~\cite{webb2019impact} identified ``elevator installers'' as one of the most impacted occupations by AI. However, he also acknowledged that this classification is an example of false positives caused by word matching methods being coarse-grained. By applying our deep-learning method instead, we learn that nearly all the tasks of the occupation ``elevator installers'' are indeed not impacted (Table~\ref{tab:elevator}): only 1 out of 20 tasks was impacted. This speaks to the robustness of a deep-learning approaches compared to a word-matching approach.

\begin{table*}[htbp]
    \centering
    \caption{
   List of the the tasks for the ``elevator installers and repairers'' occupation, and  whether each task  is impacted or not based on AII. A task is impacted when there is a patent with a similarity exceeding the 90\textsuperscript{th} percentile threshold.}
    \label{tab:elevator}
    \scalebox{0.9}{
    \begin{tabular}{|p{12cm}|c|}
        \hline
        \textbf{Task} & \textbf{Is impacted?}  \\
        \hline
        Locate malfunctions in brakes, motors, switches, and signal and control systems, using test equipment. & Yes \\
        \hdashline
        Assemble, install, repair, and maintain elevators, escalators, moving sidewalks, and dumbwaiters, using hand and power tools, and testing devices such as test lamps, ammeters, and voltmeters. & No \\
        \hdashline
        Test newly installed equipment to ensure that it meets specifications, such as stopping at floors for set amounts of time. & No \\
        \hdashline
        Check that safety regulations and building codes are met, and complete service reports verifying conformance to standards. & No \\
        \hdashline
        Connect electrical wiring to control panels and electric motors. & No \\
        \hdashline
        Adjust safety controls, counterweights, door mechanisms, and components such as valves, ratchets, seals, and brake linings. & No \\
        \hdashline
        Read and interpret blueprints to determine the layout of system components, frameworks, and foundations, and to select installation equipment. & No \\
        \hdashline
        Inspect wiring connections, control panel hookups, door installations, and alignments and clearances of cars and hoistways to ensure that equipment will operate properly. & No \\
        \hdashline
        Disassemble defective units, and repair or replace parts such as locks, gears, cables, and electric wiring. & No \\
        \hdashline
        Maintain log books that detail all repairs and checks performed. & No \\
        \hdashline
        Participate in additional training to keep skills up to date. & No \\
        \hdashline
        Attach guide shoes and rollers to minimize the lateral motion of cars as they travel through shafts. & No \\
        \hdashline
        Connect car frames to counterweights, using steel cables. & No \\
        \hdashline
        Bolt or weld steel rails to the walls of shafts to guide elevators, working from scaffolding or platforms. & No \\
        \hdashline
        Assemble elevator cars, installing each car's platform, walls, and doors. & No \\
        \hdashline
        Install outer doors and door frames at elevator entrances on each floor of a structure. & No \\
        \hdashline
        Install electrical wires and controls by attaching conduit along shaft walls from floor to floor and pulling plastic-covered wires through the conduit. & No \\
        \hdashline
        Cut prefabricated sections of framework, rails, and other components to specified dimensions. & No \\
        \hdashline
        Operate elevators to determine power demands, and test power consumption to detect overload factors. & No \\
        \hdashline
        Assemble electrically powered stairs, steel frameworks, and tracks, and install associated motors and electrical wiring. & No \\
        \hline
    \end{tabular}
    }
\end{table*}
\clearpage

\noindent
\textbf{Word-matching methods: Potential false negatives.} In addition to, at times, matching incorrectly tasks and patents (false positives), word matching methods may also fail to return genuine matches (false negatives). For example, Webb's paper reported a variety of tasks for which no patent was found (Table~\ref{tab:task_patent_verb_noun}). However, by applying AII to those tasks, we instead found that there are indeed matching patents. This discrepancy likely arises from word-matching methods discarding multiple verb-noun pairs, and overlooking details in patent abstracts.

\begin{table*}[htbp]
    \centering
    \caption{List of tasks with zero AI Exposure Score, indicating no associated patents~\cite{webb2019impact}. In contrast, AII identifies relevant patents for those tasks. }
    \label{tab:task_patent_verb_noun}
    \scalebox{0.7}{
        \begin{tabular}{|p{5cm}|c|c|p{10cm}|c|}
        \hline
        Task & Extracted pairs & AI exposure score & Most Similar Patent Abstract & Similarity \\
        \hline
        Document and maintain records of precision agriculture information. & (maintain, record) & 0.000 & A method and system for predicting soil and/or plant condition in precision agriculture with a classification of measurement data for providing an assignment of a measurement parcel to classes of interest. The assignment is used for providing action recommendations, particularly in real time or close to real time, to a farmer and/or to an agricultural device based on acquired measurement data, particularly remote sensing data, and wherein a classification model is trained by a machine learning algorithm, e.g. relying on deep learning for supervised and/or unsupervised learning, and is potentially continuously refined and adapted thanks to a feedback procedure. & 0.790 \\
        \hdashline
        Apply precision agriculture information to specifically reduce the negative environmental impacts of farming practices. & (apply, information) & 0.000 & A method and system for predicting soil and/or plant condition in precision agriculture with a classification of measurement data for providing an assignment of a measurement parcel to classes of interest. The assignment is used for providing action recommendations, particularly in real time or close to real time, to a farmer and/or to an agricultural device based on acquired measurement data, particularly remote sensing data, and wherein a classification model is trained by a machine learning algorithm, e.g. relying on deep learning for supervised and/or unsupervised learning, and is potentially continuously refined and adapted thanks to a feedback procedure. & 0.813   \\
        \hdashline
        Install, calibrate, or maintain sensors, mechanical controls, GPS-based vehicle guidance systems, or computer settings. & (maintain, sensor) & 0.000 & A vehicle computing system validates location data received from a Global Navigation Satellite System receiver with other sensor data. In one embodiment, the system calculates velocities with the location data and the other sensor data. The system generates a probabilistic model for velocity with a velocity calculated with location data and variance associated with the location data. The system determines a confidence score by applying the probabilistic model to one or more of the velocities calculated with other sensor data. In another embodiment, the system implements a machine learning model that considers features extracted from the sensor data. The system generates a feature vector for the location data and determines a confidence score for the location data by applying the machine learning model to the feature vector. Based on the confidence score, the system can validate the location data. The validated location data is useful for navigation and map updates. & 0.817 \\
        \hline
        \end{tabular}
    }
\end{table*}
\clearpage

\clearpage
\section*{Most- and Least-Impacted Occupations and Industry Sections}
\label{appendix:10-years}

\subsection*{Between 2010 and 2020}
We computed the AII measure on occupations and industry sectors in a larger time window between 2010 and 2020. Similarly, we observed that the most impacted occupations come primarily from healthcare, information technology, and manufacturing, whereas the least impacted ones come from finance and insurance, education, and construction.

\begin{table}[htbp]
    \centering
    \caption{20 most- and least-impacted occupations ranked by the AII (Artificial Intelligence Impact) measure using AI patents from 2010 to 2020. For the 20 most- and least-impacted occupations, only 1 entry differs for each. This is because patents in 2015-2020 account for 96\% of the total patents in 2010-2020.}
    \label{tab:job_aii_10y}
    \scalebox{0.8}{
    \begin{tabular}{|l|p{4cm}|p{4cm}|}
    
    \hline
    \textbf{Rank} & \textbf{Most-impacted} & \textbf{Least-impacted} \\
    \hline
    1 & Orthodontists & Pile-Driver Operators \\
    \hdashline
    2 & Cardiovascular Technologists and Technicians & Graders and Sorters, Agricultural Products \\
    \hdashline
    3 & Medical Records and Health Information Technicians & Floor Sanders and Finishers \\
    \hdashline
    4 & Multi-Media Artists and Animators & Aircraft Cargo Handling Supervisors \\
    \hdashline
    5 & Electro-Mechanical Technicians & Insurance Appraisers, Auto Damage \\
    \hdashline
    6 & Magnetic Resonance Imaging Technologists & Insurance Underwriters \\
    \hdashline
    7 & Nuclear Medicine Technologists & Reinforcing Iron and Rebar Workers \\
    \hdashline
    8 & Software Developers, Applications & Farm Labor Contractors \\
    \hdashline
    9 & Industrial Truck and Tractor Operators & Water and Liquid Waste Treatment Plant and System Operators \\
    \hdashline
    10 & Numerical Tool and Process Control Programmers & Animal Scientists \\
    \hdashline
    11 & Sound Engineering Technicians & Brokerage Clerks \\
    \hdashline
    12 & Computer Programmers & Insulation Workers, Floor, Ceiling, and Wall \\
    \hdashline
    13 & Life, Physical, and Social Science Technicians, All Other & Locomotive Firers \\
    \hdashline
    14 & Earth Drillers, Except Oil and Gas & Management Analysts \\
    \hdashline
    15 & Medical Transcriptionists & Podiatrists \\
    \hdashline
    16 & Airline Pilots, Copilots, and Flight Engineers & Cooks, Short Order \\
    \hdashline
    17 & Commercial Pilots & Shipping, Receiving, and Traffic Clerks \\
    \hdashline
    18 & Physical Scientists, All Other & Helpers--Painters, Paperhangers, Plasterers, and Stucco Masons \\
    \hdashline
    19 & Computer-Controlled Machine Tool Operators, Metal and Plastic & Team Assemblers \\
    \hdashline
    20 & Biomedical Engineers & Ambulance Drivers and Attendants, Except Emergency Medical Technicians \\
    \hline
    \end{tabular}
    }
\end{table}

\begin{figure}[htbp]
    \centering
    \includegraphics[width=\linewidth]{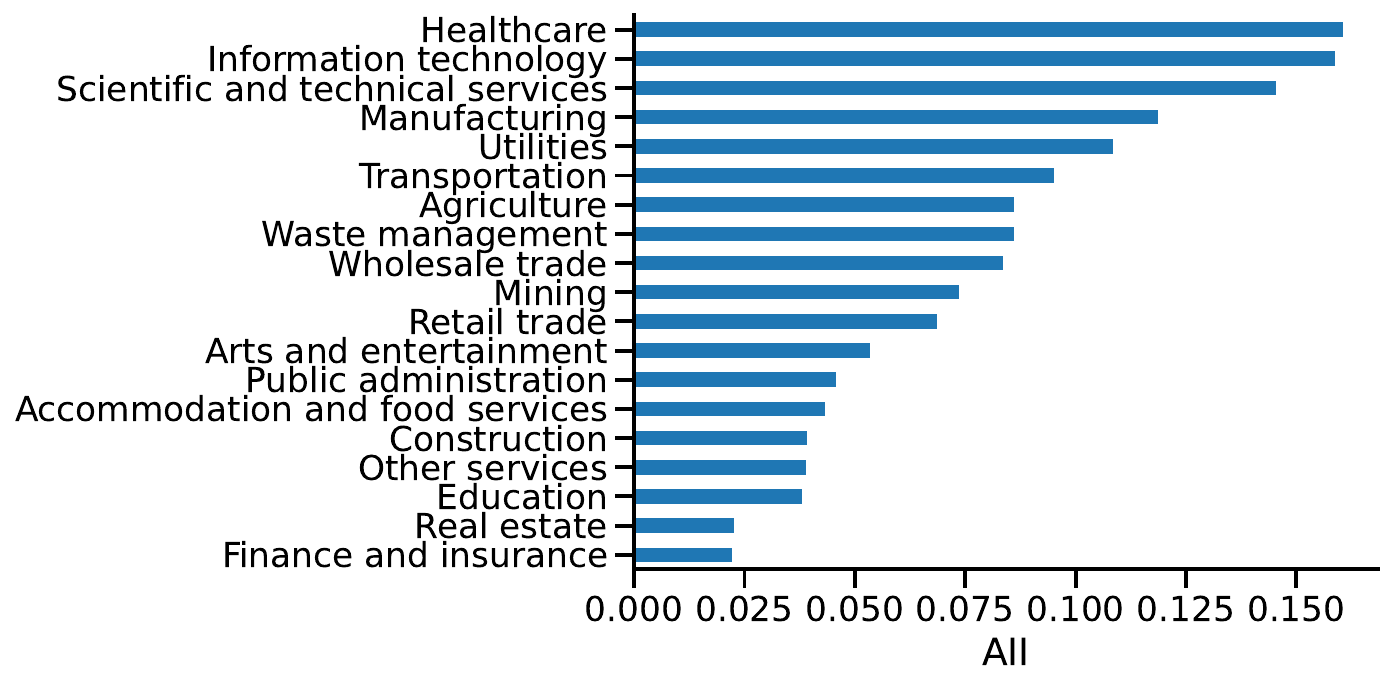}
    \caption{Overall AI impact on industry sectors from 2010-2020, ranked from highest to lowest sector-level AII scores. A sector-level AII score represents the mean AII across all occupations within that sector. The only difference from the 2015-2020 version is the ordering of other services and construction sectors.}
    \label{fig:sector-10y}
\end{figure}

\subsection*{Between 2010 and 2022}
\label{appendix:upto2022}

We computed the AII measure on occupations and industry sectors in a larger time window between 2010 and 2022. In this extended timeframe, we again observed that the most impacted occupations primarily come from healthcare, information technology, and manufacturing. Additionally, we noted an increasing impact of AI in scientific occupations.

\begin{table}[htbp]
    \centering
    \caption{20 most- and least-impacted occupations ranked by the AII (Artificial Intelligence Impact) measure using AI patents from 2010 to 2022.}
    \label{tab:job_aii_2022}
    \scalebox{0.8}{
    \begin{tabular}{|l|p{4cm}|p{4cm}|}
    
    \hline
    \textbf{Rank} & \textbf{Most-impacted} & \textbf{Least-impacted} \\
    \hline
    1 & Cardiovascular Technologists and Technicians & Pile-Driver Operators \\
    \hdashline
    2 & Sound Engineering Technicians & Graders and Sorters, Agricultural Products \\
    \hdashline
    3 & Nuclear Medicine Technologists & Floor Sanders and Finishers \\
    \hdashline
    4 & Magnetic Resonance Imaging Technologists & Aircraft Cargo Handling Supervisors \\
    \hdashline
    5 & Air Traffic Controllers & Insurance Underwriters \\
    \hdashline
    6 & Orthodontists & Reinforcing Iron and Rebar Workers \\
    \hdashline
    7 & Electro-Mechanical Technicians & Farm Labor Contractors \\
    \hdashline
    8 & Power Distributors and Dispatchers & Rock Splitters, Quarry \\
    \hdashline
    9 & Industrial Truck and Tractor Operators & Brokerage Clerks \\
    \hdashline
    10 & Police, Fire, and Ambulance Dispatchers & Locomotive Firers \\
    \hdashline
    11 & Security Guards & Management Analysts \\
    \hdashline
    12 & Physical Scientists, All Other & Podiatrists \\
    \hdashline
    13 & Machinists & Cooks, Short Order \\
    \hdashline
    14 & Atmospheric and Space Scientists & Shipping, Receiving, and Traffic Clerks \\
    \hdashline
    15 & Computer-Controlled Machine Tool Operators, Metal and Plastic & Proofreaders and Copy Markers \\
    \hdashline
    16 & Medical Records and Health Information Technicians & Helpers--Painters, Paperhangers, Plasterers, and Stucco Masons \\
    \hdashline
    17 & Textile Knitting and Weaving Machine Setters, Operators, and Tenders & Team Assemblers \\
    \hdashline
    18 & Radio Operators & Political Scientists \\
    \hdashline
    19 & Medical Transcriptionists & Paralegals and Legal Assistants \\
    \hdashline
    20 & Physician Assistants & Telemarketers \\
    \hline
    \end{tabular}
    }
\end{table}

\begin{figure}[htbp]
    \centering
    \includegraphics[width=\linewidth]{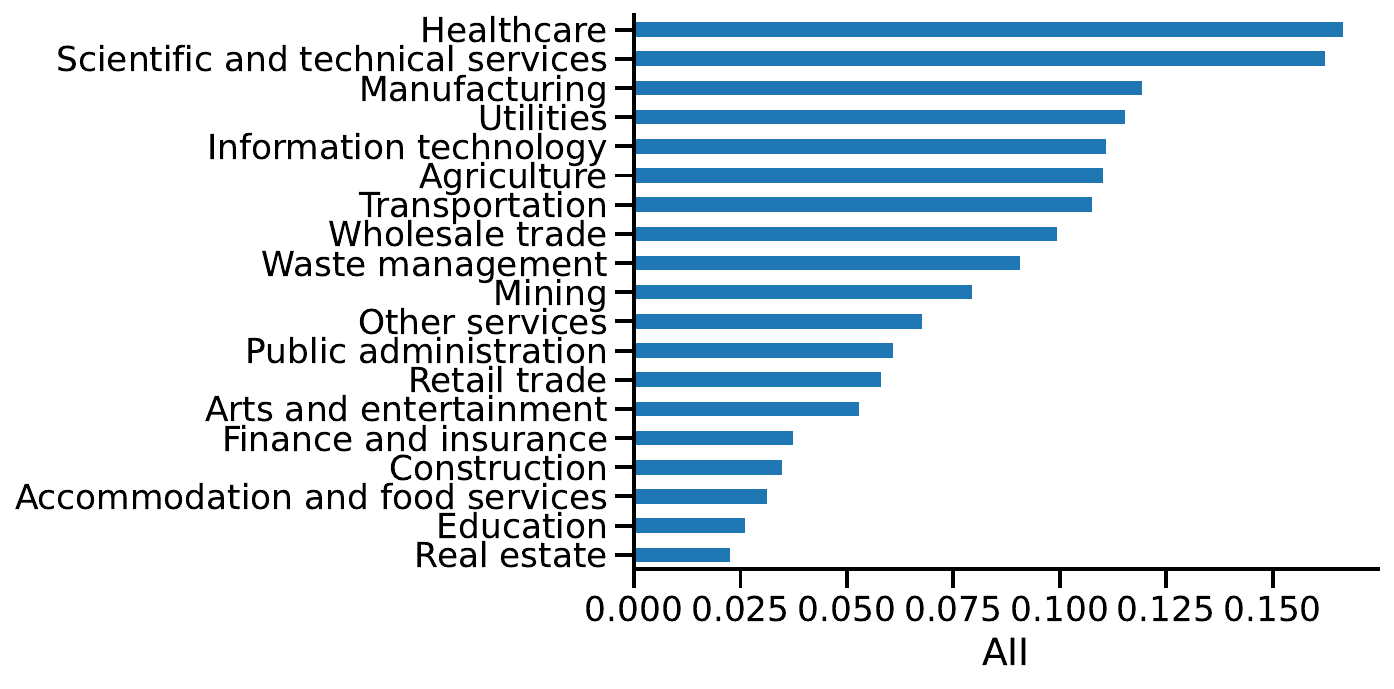}
    \caption{Overall AI impact on industry sectors from 2010-2022, ranked from highest to lowest sector-level AII scores. A sector-level AII score represents the mean AII across all occupations within that sector.}
    \label{fig:sector-2022}
\end{figure}

\clearpage
\subsection*{Using patents from the U.S., China, Japan, and Korea}
\label{appendix:patents-extended}

We  collected all 153,854 patents from China, Japan, and South Korea written in English from Google Patents Public Data during the period of study. The AI patents from these countries and the U.S. make up 81\% of the total published AI patents.
 
\begin{figure}[t!]
    \centering
    \includegraphics[width=.5\textwidth]{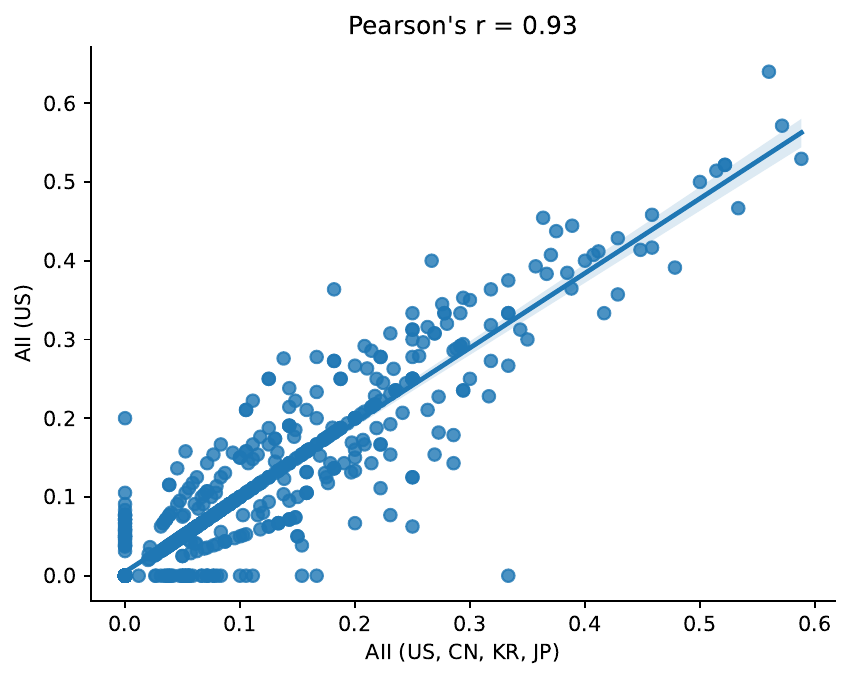}
    \caption{Correlation between AII calculated with only US patents and with patents from US, China, and Japan.}
    \label{fig:aii-non-us}
\end{figure}

We created two sets that contain patents from: \emph{(1)} U.S. only; and \emph{(2)} U.S., China, Japan, and Korea. As seen in Figure~\ref{fig:aii-non-us}, the AII scores computed on the two sets of patents have a correlation of $r = 0.93$. With the newly added patents from China, Japan, and Korea, additional occupations potentially impacted include  ``bakers'', ``solar photovoltaic installers'', and ``dredge operators'', predominantly affected by patents originating from China. 

\subsection*{Job Vacancy Rates by Sector \emph{vs.} sector-level AII}
\setcurrentname{Job Vacancy Rates by Sector}\phantomsection\label{appendix:job-vacancy-sector}

\begin{figure}[h!]
    \centering
    \includegraphics[width=.4\textwidth]{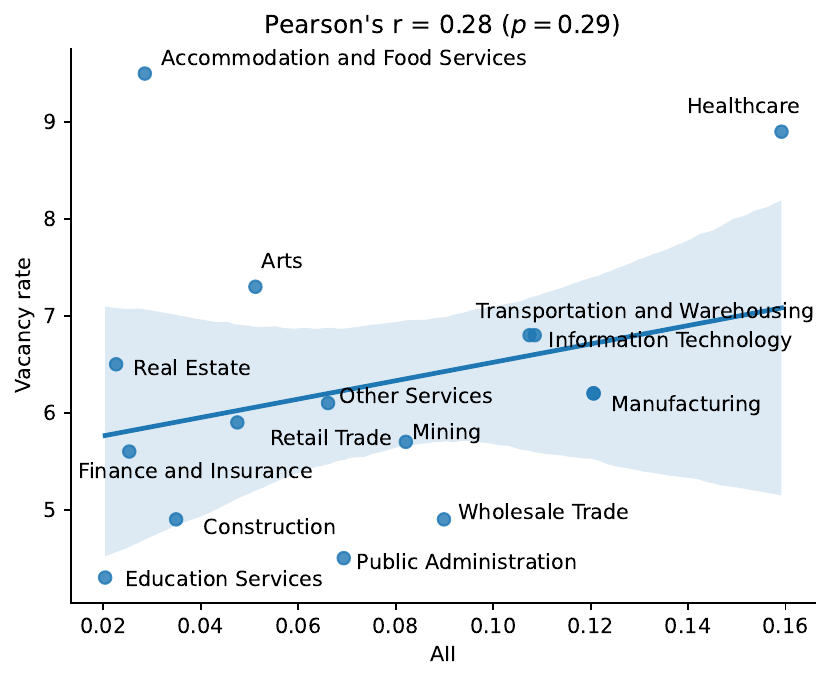} 
    \caption{Job vacancy rates by sector vs. sector-level AII with all industry sectors.}
    \label{fig:vacancy-sector-outlier}
\end{figure}

AII at sector level and vacancy rates are positively correlated, with a Pearson's correlation coefficient of $r = 0.28$ ($ p = 0.29$) but not statistically significant (Figure~\ref{fig:vacancy-sector-outlier}).

\clearpage
\section*{Impact of Robots and Software}
\setcurrentname{Impact of Robots and Software}\phantomsection\label{appendix:wage}

We investigated the relationship between the adapted AII score for robots and software and changes in wages and employment between 1980 and 2010. We did so by following Webb's methodology~\cite{webb2019impact}, and by using census data and patents related to robots or software.

\subsection*{Data}

For robot-related patents, we filtered patents based on whether the keywords ``robot'' or ``manipulat'' appeared in the patents' titles and abstracts. Additionally, we ensured the resulting set of patents did not have Cooperative Patent Classification (CPC) codes A61 (medical or veterinary science; hygiene) or B01 (physical or chemical processes or apparatus in general), following Webb's methodology~\cite{webb2019impact}. For software-related patents, our filter included any patent with the keywords ``software'', ``computer'', or ``program'', while excluding any that mentioned ``chip'', ``semiconductor'', ``bus'', ``circuity'', or ``circuitry''.

To compute the adapted AII score for either robots or software, we applied the same formula (Equation~\ref{eq:aii}) used to compute the impact of AI at the task level but replaced the set of patents with those associated either with robots or software. To compute the impact of robots or software at the occupation level, we used the same formula (Equation~\ref{eq:weighted_job}) because it depends on the number of tasks that are impacted rather than the patents themselves.

For changes in wages and employment, we used individual-level microdata from the US Census between 1960 and 2000 and from the ACS between 2000 and 2018, both of which were provided in the Integrated Public Use Microdata
Series (IPUMS)~\cite{ruggles2024ipums}. We restricted the analysis to individuals aged between 18 and 65 who were employed and engaged in some form of work.\footnote{We used the field WORKEDYR from the IPUMS data, which indicates whether the person had worked at all for profit, pay, or as an unpaid family worker during the previous year. For the census samples, the reference period is the previous calendar year; for the ACS and the PRCS, the reference period is the preceding 12 months.} We calculated the average wages and the proportion of hours worked within each industry-occupation pair (referred to as an industry-occupation cell). We used a number of additional census control variables including age, gender, level of education, and offshorability (i.e., the degree to which an occupation requires either direct interpersonal interaction or proximity to a specific work location). The measure of offshorability was developed by~\cite{firpo2011occupational} and standardized by~\cite{autor2013growth}. For all the subsequent analyses, we used the IPUMS ``occ1990'' occupational classification and ``ind1990'' industry classification.

To ensure that the relationship between the AII scores for robots and software and changes in employment and wages is measured proportionately to the workforce size, we applied a number of controls following Webb's methodology~\cite{webb2019impact}. First, we applied the IPUMS survey's individual weight (PERWT) adjusted by the proportion of full-time work in each industry and occupation combination, resulting into a ``labor-supply weight''. This adjustment yields the total number of full-time-equivalent (FTE) employees in each industry-occupation pair. Additionally, we introduced a ``demographic-adjusted'' labor-supply weight to reflect demographic shifts since 1980. This involved categorizing demographic groups by gender, race (black, white, other), level of education (less than high school, high school graduate, some college, bachelor's degree), and age group (in five-year intervals). The adjusted weight for 2010 was then computed by maintaining each group's proportionate weight from 1980 (that is, adjusting each data point by the 1980 to 2010 weight ratio for its demographic group). For wage calculations, we determined the real weekly wages (in 2016 dollars) for 1980 and 2010, focusing on full-time-full-year (FTFY, over 35 hours/week and 40 weeks/year) employees, applying a 98\% winsorization to control for extreme values in the earnings data annually. We then aggregated the census data into industry-occupation-year groups using the ``ind1990'' and ``occ1990'' classifications from IPUMS. In our subsequent regression analyses, the education variable was adjusted according to the labor-supply-weighted average years of education for each industry-occupation cell in 1980, categorized into terciles.

\subsection*{Regression Model}

To measure the relationship between the adapted AII score for robots and software and changes in employment and wages, we estimated variations of the following regression:
\begin{equation}
    \Delta y_{o,i,t} = \alpha_i + \beta x_o + \gamma Z_o + \varepsilon_{o,i,t}
    \label{eq:regression_formula}
\end{equation}
The unit of observation is an occupation-industry-year cell, such as welders in auto manufacturing in 1980, with $o$ denoting occupation, $i$ industry, and $t$ year. The dependent variable is the difference between 1980 and 2010 of an outcome variable of interest (i.e., employment and wages). On the right-hand side (Equation~\ref{eq:regression_formula}), we included industry fixed effects $\alpha_i$; the exposure of the occupation to robots or software $x_o$; and the vector of controls $Z_o$ contains occupation-level variables such as terciles of average years of education and offshorability.

To measure the change in employment, we used the DHS changes, also known as arc percentage change or percent change relative to the midpoint. DHS is a symmetric measure of the growth rate defined as the difference between two values divided by their average~\cite{davis1996job}. This results in a second-order approximation of the log change for growth rates near zero; values are restricted to being between -2 and 2, with -2 and 2 representing exit and entry respectively.

To measure the change in wages, we used the log change in real weekly wages for full-time and full-year workers in each industry-occupation cell. For the adapted AII scores, we transformed the raw scores to be in employment-weighted percentiles. Thus, a score of 90 means that 10\% of workers work in occupations with a higher exposure. To construct the industry-occupation cell and account for industry fixed effects, we used the IPUMS ``ind1990'' industry classification code. Finally, the sample used to measure the relationship between exposure to robots and changes in employment and wages is restricted to industries within the manufacturing sector.

\subsection*{Results}
The models show that moving from the 25\textsuperscript{th} to the 75\textsuperscript{th} percentile of exposure to robots is associated with a decline in wages between 2 and 4\% (Table~\ref{tab:eval-wage-robots}), depending on the specification, and a varying effect in industry employment share of between -9 and 18\% (Table~\ref{tab:eval-emp-robots}). Similarly, moving from the 25\textsuperscript{th} to the 75\textsuperscript{th} percentile of exposure to software is associated with a decline in wages of between 4 and 7\% (Table~\ref{tab:eval-wage-software}), and a decline in within-industry employment shares of between 3 and 10\% (Table~\ref{tab:eval-emp-software}). Recall that these are within-industry effects. Therefore, these results do not simply say that manufacturing jobs are exposed to robots, and manufacturing has (for other reasons) declined. Rather, they show that within each manufacturing industry, the particular occupations exposed to robots have declined much more than those that are not exposed.

However, these relationships might be influenced by external factors such as offshorability, industry effects, educational levels, and wage polarization. Given the emergence of offshoring as a significant trend between 1980 and 2010, we incorporated it as a control variable in our analysis. Additionally, we anticipated that industry-specific effects would capture variations stemming from trade dynamics and evolving consumer preferences impacting product demand. Another consideration was the significant demographic and skill shifts observed between 1980 and 2010. It is plausible therefore that the workforce's educational enhancement, particularly the increase in highly skilled individuals, altered the labor supply dynamics, diminishing the availability of low-skilled workers who are most vulnerable to automation. This shift could mislead us into attributing the decrease in low-skilled labor demand to automation, overlooking the actual supply reduction. Finally, the potential of wage polarization, not directly linked to automation, is another factor to control for because previous economic studies have shown a decline in middle-skill workers' wages within the timeframe of our study, contrasting with a rise in top-tier salaries \cite{michaels2014has}. Despite accounting for these factors, the relationship between exposure to robots or software and changes in employment or wages remained statistically significant.

\begin{table*}[t]
    \centering
    \caption{Change in wages vs. exposure to robots, 1980-2010. Each observation is an occupation-industry cell. Dependent variable is 100x change in log wage between 1980 and 2010, winsorized at the top and bottom 1\%. Education variables are terciles of average years of education for occupation-industry cells in 1980. Wages are cells' mean weekly wage for full-time, full-year workers in 1980. Offshorability is an occupation-level measure from Autor and Dorn (2013). Observations are weighted by cell's labor supply, averaged between 1980 and 2010.}
    \label{tab:eval-wage-robots}
    \scalebox{0.8}{
    \begin{tabular}{@{\extracolsep{5pt}}lcccccc}
    \\[-1.8ex]\hline\hline \\
    [-1.8ex] & (1) & (2) & (3) & (4) & (5) & (6) \\
    \hline \\[-1.8ex]
    AII & -0.076$^{***}$ & -0.076$^{***}$ & -0.076$^{***}$ & -0.035$^{*}$ & -0.042$^{**}$ & \\
    & (0.018) & (0.018) & (0.018) & (0.018) & (0.018) & \\
    Offshorability & & & -2.029$^{***}$ & -0.693$^{}$ & 5.077$^{***}$ & 5.162$^{***}$ \\
    & & & (0.701) & (0.713) & (0.730) & (0.729) \\
    Medium education & & & & 5.271$^{***}$ & 9.529$^{***}$ & 9.756$^{***}$ \\
    & & & & (0.962) & (0.941) & (0.937) \\
    High education & & & & 8.732$^{***}$ & 27.022$^{***}$ & 27.503$^{***}$ \\
    & & & & (0.984) & (1.276) & (1.261) \\
    Wage & & & & & -0.056$^{***}$ & -0.056$^{***}$ \\
    & & & & & (0.003) & (0.003) \\
    Wage squared & & & & & 0.000$^{***}$ & 0.000$^{***}$ \\
    & & & & & (0.000) & (0.000) \\
    \hline \\[-1.8ex]
    Adjusted $R^2$ & 0.003 & 0.036 & 0.037 & 0.049 & 0.126 & 0.125 \\
    Industry FEs & & \checkmark & \checkmark & \checkmark & \checkmark & \checkmark \\
    Observations & 5957 & 5957 & 5957 & 5957 & 5957 & 5957 \\
    \hline
    \hline \\[-1.8ex]
    \multicolumn{7}{p{.8\textwidth}}{Notes: $^{*}$p$<$0.1; $^{**}$p$<$0.05; $^{***}$p$<$0.01}
    \end{tabular}
    }
\end{table*}
\begin{table*}[t]
    \centering
    \caption{Change in employment vs. exposure to robots, 1980-2010. Each observation is an occupation-industry cell. Dependent variable is 100x DHS change of a cell's share of overall employment between 1980 and 2010, winsorized at the top and bottom 1\%. Education variables are terciles of average years of education for occupation-industry cells in 1980. Wages are cells' mean weekly wage for full-time, full-year workers in 1980. Offshorability is an occupation-level measure from Autor and Dorn (2013). Observations are weighted by cell's labor supply, averaged between 1980 and 2010.}
    \label{tab:eval-emp-robots}
    \scalebox{0.8}{
    \begin{tabular}{@{\extracolsep{5pt}}lcccccc}
    \\[-1.8ex]\hline\hline \\
    [-1.8ex] & (1) & (2) & (3) & (4) & (5) & (6) \\
    \hline \\[-1.8ex]
    AII & 0.360$^{***}$ & 0.373$^{***}$ & 0.360$^{***}$ & -0.124$^{}$ & -0.188$^{**}$ & \\
    & (0.064) & (0.063) & (0.063) & (0.080) & (0.079) & \\
    Offshorability & & & 4.639$^{*}$ & 0.571$^{}$ & 15.211$^{***}$ & 15.233$^{***}$ \\
    & & & (2.514) & (2.535) & (2.651) & (2.652) \\
    Medium Education & & & & -29.584$^{***}$ & -20.301$^{***}$ & -15.843$^{***}$ \\
    & & & & (3.916) & (3.885) & (3.408) \\
    High Education & & & & -40.941$^{***}$ & 6.256$^{}$ & 12.004$^{***}$ \\
    & & & & (4.296) & (5.180) & (4.588) \\
    Wage & & & & & -0.113$^{***}$ & -0.113$^{***}$ \\
    & & & & & (0.012) & (0.012) \\
    Wage squared & & & & & 0.000$^{***}$ & 0.000$^{***}$ \\
    & & & & & (0.000) & (0.000) \\
    \hline \\[-1.8ex] 
    Adjusted $R^2$ & 0.005 & 0.067 & 0.067 & 0.082 & 0.119 & 0.118 \\
    Industry FEs & & \checkmark & \checkmark & \checkmark & \checkmark & \checkmark \\
    Observations & 5957 & 5957 & 5957 & 5957 & 5957 & 5957 \\
    \hline
    \hline \\[-1.8ex]
    \textit{Note:} & \multicolumn{6}{r}{$^{*}$p$<$0.1; $^{**}$p$<$0.05; $^{***}$p$<$0.01} \\
    \end{tabular}
    }
\end{table*}

\begin{table*}[t]
    \centering
    \caption{Change in wages vs. exposure to software, 1980-2010. Each observation is an occupation-industry cell. Dependent variable is 100x change in log wage between 1980 and 2010, winsorized at the top and bottom 1\%. Education variables are terciles of average years of education for occupation-industry cells in 1980. Wages are cells' mean weekly wage for full-time, full-year workers in 1980. Offshorability is an occupation-level measure from Autor and Dorn (2013). Observations are weighted by cell's labor supply, averaged between 1980 and 2010.}
    \label{tab:eval-wage-software}
    \scalebox{0.8}{
    \begin{tabular}
    {@{\extracolsep{5pt}}lcccccc}
    \\[-1.8ex]\hline\hline \\
    [-1.8ex] & (1) & (2) & (3) & (4) & (5) & (6) \\
    \hline \\[-1.8ex]
    AII & -0.141$^{***}$ & -0.131$^{***}$ & -0.135$^{***}$ & -0.068$^{**}$ & -0.077$^{***}$ & \\
    & (0.030) & (0.030) & (0.030) & (0.031) & (0.029) & \\
    Offshorability & & & -3.498$^{***}$ & 0.105$^{}$ & 14.785$^{***}$ & 15.036$^{***}$ \\
    & & & (1.106) & (1.150) & (1.133) & (1.129) \\
    Medium education & & & & 9.695$^{***}$ & 24.125$^{***}$ & 24.380$^{***}$ \\
    & & & & (1.678) & (1.617) & (1.614) \\
    High education & & & & 19.347$^{***}$ & 73.521$^{***}$ & 74.252$^{***}$ \\
    & & & & (1.743) & (2.051) & (2.033) \\
    Wage & & & & & -0.170$^{***}$ & -0.171$^{***}$ \\
    & & & & & (0.005) & (0.005) \\
    Wage squared & & & & & 0.000$^{***}$ & 0.000$^{***}$ \\
    & & & & & (0.000) & (0.000) \\
    \hline \\[-1.8ex]
    Adjusted $R^2$ & 0.001 & 0.028 & 0.028 & 0.035 & 0.137 & 0.136 \\
    Industry FEs & & \checkmark & \checkmark & \checkmark & \checkmark & \checkmark \\
    Observations & 18724 & 18724 & 18724 & 18724 & 18724 & 18724 \\
    \hline
    \hline \\[-1.8ex]
    \textit{Note:} & \multicolumn{6}{r}{$^{*}$p$<$0.1; $^{**}$p$<$0.05; $^{***}$p$<$0.01} \\
    \end{tabular}
    }
\end{table*}
\begin{table*}[t]
    \centering
    \caption{Change in employment vs. exposure to software, 1980-2010. Each observation is an occupation-industry cell. Dependent variable is 100x DHS change of a cell's share of overall employment between 1980 and 2010, winsorized at the top and bottom 1\%. Education variables are terciles of average years of education for occupation-industry cells in 1980. Wages are cells' mean weekly wage for full-time, full-year workers in 1980. Offshorability is an occupation-level measure from Autor and Dorn (2013). Observations are weighted by cell's labor supply, averaged between 1980 and 2010.}
    \label{tab:eval-emp-software}
    \scalebox{0.8}{
    \begin{tabular}{@{\extracolsep{5pt}}lcccccc}
    \\[-1.8ex]\hline\hline \\
    [-1.8ex] & (1) & (2) & (3) & (4) & (5) & (6) \\
    \hline \\[-1.8ex]
    AII & -0.126$^{***}$ & -0.063$^{}$ & -0.057$^{}$ & -0.196$^{***}$ & -0.195$^{***}$ & \\
    & (0.042) & (0.040) & (0.040) & (0.040) & (0.040) & \\
    Offshorability & & & 5.616$^{***}$ & -2.606$^{*}$ & 4.628$^{***}$ & 5.263$^{***}$ \\
    & & & (1.452) & (1.500) & (1.550) & (1.545) \\
    Medium education & & & & -29.554$^{***}$ & -22.455$^{***}$ & -21.809$^{***}$ \\
    & & & & (2.188) & (2.212) & (2.209) \\
    High education & & & & -43.686$^{***}$ & -17.769$^{***}$ & -15.918$^{***}$ \\
    & & & & (2.273) & (2.807) & (2.783) \\
    Wage & & & & & -0.093$^{***}$ & -0.094$^{***}$ \\
    & & & & & (0.007) & (0.007) \\
    Wage squared & & & & & 0.000$^{***}$ & 0.000$^{***}$ \\
    & & & & & (0.000) & (0.000) \\
    \hline \\[-1.8ex]
    Adjusted $R^2$ & 0.000 & 0.114 & 0.115 & 0.132 & 0.146 & 0.145 \\
    Industry FEs & & \checkmark & \checkmark & \checkmark & \checkmark & \checkmark \\
    Observations & 18724 & 18724 & 18724 & 18724 & 18724 & 18724 \\
    \hline
    \hline \\[-1.8ex]
    \textit{Note:} & \multicolumn{6}{r}{$^{*}$p$<$0.1; $^{**}$p$<$0.05; $^{***}$p$<$0.01} \\
    \end{tabular}
    }
\end{table*}

\clearpage

\section*{Beyond Automation: Measuring Augmentation}
\setcurrentname{Beyond Automation: Measuring Augmentation}\phantomsection\label{appendix:automation-augmentation}

\begin{figure}[t!]
    \centering
    \includegraphics[width=.45\textwidth]{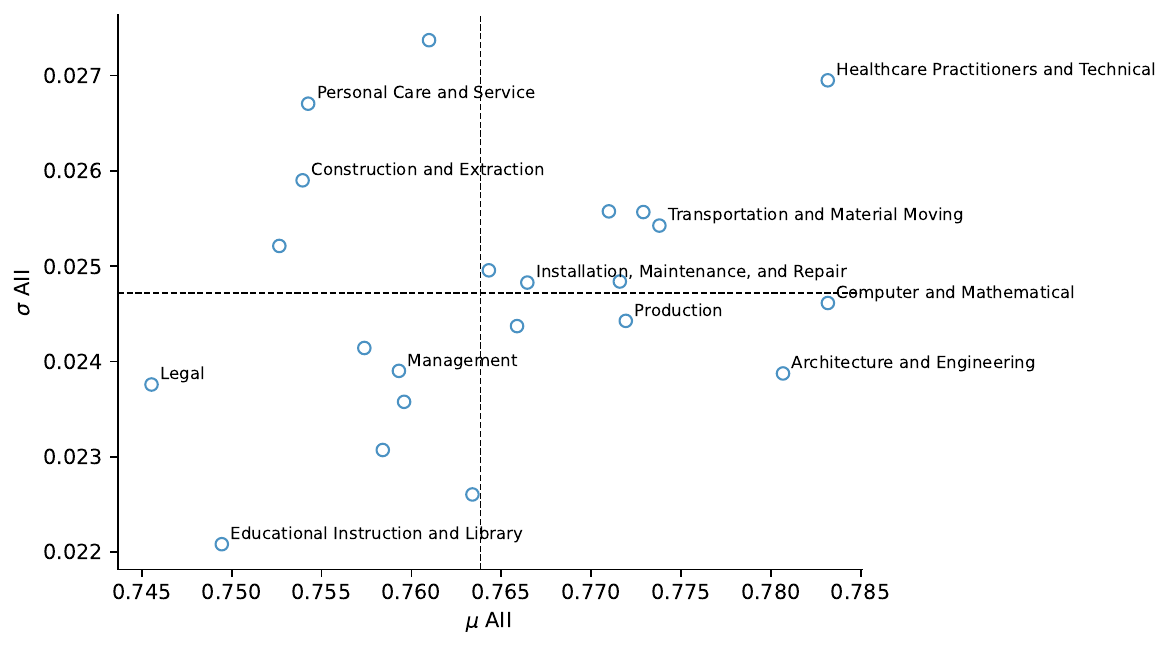}
    \caption{Automation \emph{vs.} augmentation potential computed using the mean and standard deviation of similarity scores as per \cite{gmyrek2023genai}  (bottom-left: not impacted; top-left: augmented; top-right: the big unknown; bottom-right: automated).}
    \label{fig:auto-aug-sim-group}
\end{figure}

\begin{figure}[t!]
    \centering
    \includegraphics[width=.45\textwidth]{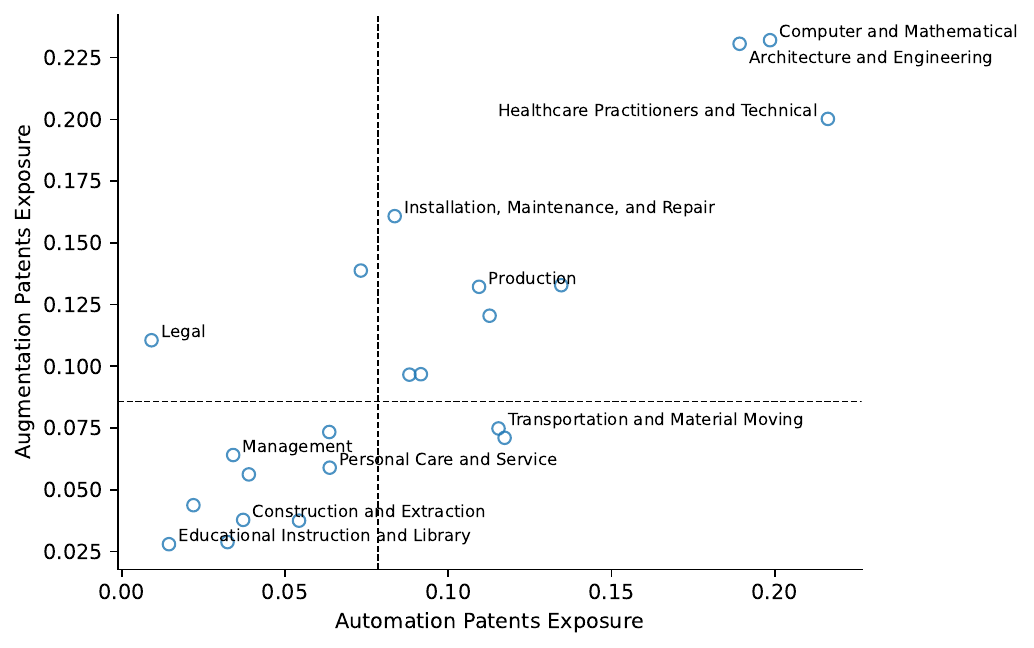}
    \caption{Automation \emph{vs.} augmentation using patent similarity to tasks and micro-titles defined in the Census Alphabetical Index of Occupations and Industries (CAI)~\cite{autor2022new}.}
    \label{fig:auto-aug-micro-titles-group}
\end{figure}

Figure~\ref{fig:auto-aug-sim-group} shows the results as per Gmyrek \emph{et al.}~\cite{gmyrek2023genai}'s method. To compare Autor \emph{et al.}'s~\cite{autor2022new} and Gmyrek \emph{et al.}'s~\cite{gmyrek2023genai} methods, for each occupation group defined by the first two digits---coarser-grained classification compared to the six digits classification---of the Standard Occupational Classification (SOC) code (\url{https://www.bls.gov/soc/socguide.htm}), we computed the average similarity value for automation and augmentation. We then took the median of the two average values and created a quadrant (automation vs. augmentation). The top left quadrant then indicates occupations that are likely to be exposed to augmentation, while the bottom right quadrant shows the exposure to automation. Figure~\ref{fig:auto-aug-micro-titles-group} shows the results as per Autor et al~\cite{autor2022new}'s method, adapted to be visually similar with that of Figure~\ref{fig:auto-aug-sim-group}.  From these two figures, we observed that, in line with our results, Gmyrek \emph{et al.}~\cite{gmyrek2023genai}'s method identified personal care (e.g., hearing aid specialist), construction (e.g., electrical and electronics repairers), and food preparation (e.g., food science technician) occupations as more exposed to augmentation than automation.

\end{document}